\documentclass[conference]{IEEEtran}
\usepackage[english]{babel}
\usepackage[utf8x]{inputenc}
\usepackage[T1]{fontenc}
\usepackage{graphicx}
\usepackage{booktabs}
\usepackage{multirow}
\makeatletter
\providecommand{\tabularnewline}{\\}
\usepackage{graphics}% for pdf, bitmapped graphics files
\usepackage{pdfpages}
\usepackage{float}
\usepackage{cite}
\usepackage{balance}
\makeatother

\begin{document}

\title{Detecting Target-Area Link-Flooding DDoS Attacks using Traffic Analysis and Supervised Learning}

%\title{Traffic Analysis exposing Vulnerability of the Crossfire Attack to Early Detection}

\author{\IEEEauthorblockN{Mostafa Rezazad$^1$, Matthias R. Brust$^2$, Mohammad Akbari$^3$, Pascal Bouvry$^2$, Ngai-Man Cheung$^1$}
\IEEEauthorblockA{$^1$Singapore University of Technology and Design, \{mostafa, ngaiman\_cheung\}@sutd.edu.sg\\
$^2$SnT, University of Luxembourg, \{matthias.brust, pascal.bouvry\}@uni.lu\\
$^3$SAP Innovation Center Singapore, mohammad.akbari@sap.com  
}}

\maketitle
\begin{abstract}

A novel class of extreme link-flooding DDoS (Distributed Denial of Service) attacks is designed to cut off entire geographical areas such as cities and even countries from the Internet by simultaneously targeting a selected set of network links. The Crossfire attack is a target-area link-flooding attack, which is orchestrated in three complex phases. The attack uses a massively distributed large-scale botnet to generate low-rate \emph{benign} traffic aiming to congest selected network links, so-called \emph{target links}. The adoption of benign traffic, while simultaneously targeting multiple network links, makes detecting the Crossfire attack a serious challenge. %Although the Crossfire attack is a potential threat to any network, we show in this paper that the adversary has also substantial obstacles in the successful attack execution, which exposes the attack to detection vulnerabilities. 
%In this paper, we show through analysis and emulation, that the adversary has substantial obstacles in successfully running the attack execution, which exposes the attack to detection vulnerabilities. 
%Our approach consists of replicating the Crossfire attack in a realistic test bed emulation and to analyze the traffic. %The traffic has been measured during the topology construction phase and attack phase, and, subsequently, the obtained data has been analyzed to expose patterns and vulnerabilities of the Crossfire attack. 
In this paper, we present analytical and emulated results showing hitherto unidentified vulnerabilities in the execution of the attack, such as a correlation between coordination of the botnet traffic and the quality of the attack, and a correlation between the attack distribution and detectability of the attack. Additionally, we identified a warm-up period due to the bot synchronization. % after the attack is launched and before the target links are overwhelmed, 
%which can be used for an early attack detection as we show in this paper. %We present our results and discuss their consequences for the design of novel approaches for early Crossfire attack detection and attack mitigation measures.
%We applied machine learning to investigate the effect of traffic distribution in concealing and detecting the Crossfire attack from available traffic data. 
%We design a prototypical Crossfire attack detector, which exploits these vulnerabilities. To show the feasibility of detection, we report results of using  different machine learning models trained in various scenarios using the link volume as the main feature set.
For attack detection, we report results of using two supervised machine learning approaches: Support Vector Machine (SVM) and Random Forest (RF) for classification of network traffic to normal and abnormal traffic, i.e, attack traffic. These machine learning models have been trained in various scenarios using the link volume as the main feature set.
\end{abstract}

\section{Introduction: The Crossfire Attack}

A novel class of extreme link-flooding DDoS (Distributed Denial of Service) attacks~\cite{xue2014towards} is the \emph{Crossfire attack}, which is designed to cut off entire geographical areas such as cities and even countries from the Internet by simultaneously targeting a selected set of network links~\cite{dimitrios1, dimitrios2}. The most intriguing property of this target-area link-flooding attack is the usage of legitimate traffic flows to achieve its devastating impact by making the attack particularly difficult to detect and, consequently, to mitigate~\cite{6547106}.

The Crossfire attack uses a complex and massively large-scale botnet for attack execution~\cite{6547106}. A botnet is a network of computers infected with malware (\emph{bots}) that can be controlled remotely. A command-and-control unit updates the bots by sending them the commands of the \emph{botmaster}, which is orchestrating the attack by executing the attack procedure. The bots direct their low-intensity flows to a large number of servers in such a devastating manner that the targeted geographical region is essentially cut off from the Internet.

The success of the attack depends highly on the network structure and how the attacker plans and initiates the attack sequence~\cite{DoS4,ramazani2016topological}. The attacker aims to find a set of target links which connects to the decoy servers such that if the target links are flooded, traffic destined to the target area is prevented from reaching its destination. Reciprocally, access from the target area to Internet services outside the target area will be cut off. For the adversary to achieve its goal, it chooses public servers either inside of the target area or nearby the target area, which can be easily found due to their availability. 
%Eventually refer to the figure I will put.
%Improve the following sentence:
The quality of the attack depends on the specific selection of servers and the resulting links to be targeted, but also on the overall network topology~\cite{brust2012clustering}.

%MERGE FROM OTHER PAPER: %The focus of this study is on the Crossfire attack. In this section, we explain the Crossfire attack in more detail, but also describe attacks, which exhibits similarities to the Crossfire attack.
%Before diving deep into the characteristics and dynamics of the Crossfire attack, in this section, we explain the fundamentals of the attack. % along with similar alternative attacks

\begin{figure}
\centering \includegraphics[width=1\linewidth]{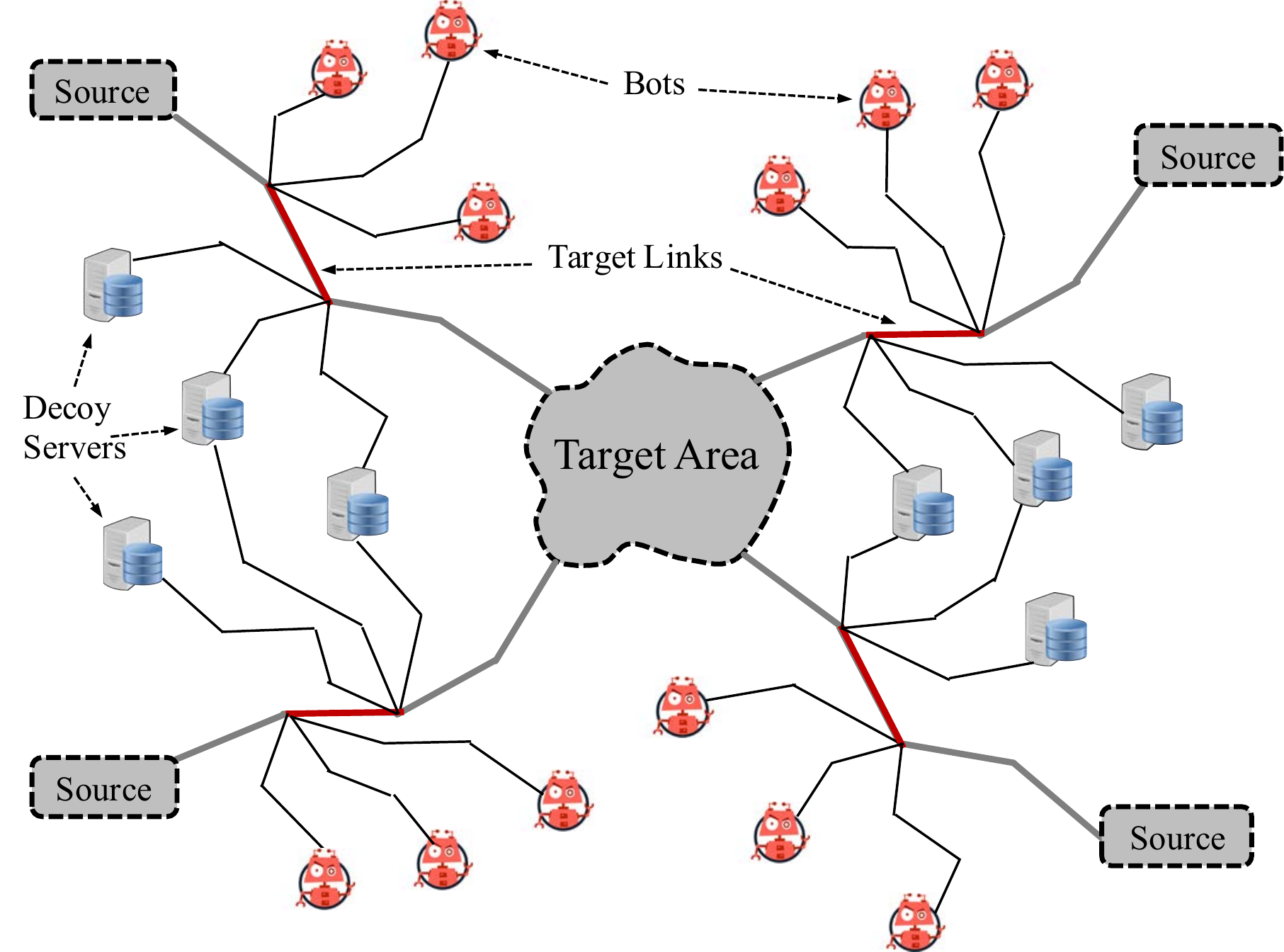}
\caption{\label{fig:crossfire}The Crossfire attack traffic flows congest a small set of selected network links using \emph{benign} low-rate flows from bots to publicly accessible servers, while degrading connectivity to the target area.}
\end{figure}

\paragraph*{The Crossfire attack}
The Crossfire attack consists of three phases: (a) the construction of the link map, (2) the selection of target links, and (c) the coordination of the botnet. While phases (a) and (b) are sequentially executed only initially, once triggered phase (c) is executed periodically. Fig.~\ref{fig:crossfire} illustrates the dynamics of the Crossfire attack.%\footnote{The detailed Crossfire attack consists of multiple sub-phases and their detailed description can be encountered in \cite{6547106}.}

\paragraph*{Link map construction}
The initial step of the Crossfire attack is the construction of the link map. The attacker crates a map of the network along the ways from the attacker's bots to the servers using \emph{traceroute}. The result of \emph{traceroute} inevitably consist of a record of different routes between the same pairs of nodes, because of network-inherent elements influencing the effective route chosen (e.g., ISP and load-balancing). Subsequently, a link map is gradually constructed, which exposes the network structure and the traffic flow behavior around the target area.

\subsubsection{Target link selection}
After the construction of the link map, the adversary evaluates the data for more stable and reliable routes to decide on its selection of the target links. 
%This selection can occur based on different criteria, e.g. how often a link appears in the routes. 
The adversary prefers disjoint routes with mostly independent target links for the attack to create the biggest impact. %The attacker can also create subsets of target links which are flooded alternatively, making the attack even more difficult to detect and to mitigate. Since we are studying the attack itself, not its formation, we assume these two phases (namely, Link Map Building and Target Link Selection) are already completed by the attacker.

\paragraph*{Bot coordination}
In the final phase of the attack, the adversary coordinates the bots to generate low-intensity traffic and to send it to the corresponding decoy server. The targeted aggregation of multiple low-intensity traffic flows on the target link ideally exhausts its capacity, hence, congesting the link. %The flows' low-intensity property represents a major challenge for existing detection methods and mitigation measures. To make detection even more complicating, the bots can be coordinated to selectively apply alternating attack behaviors. %, e.g. sub-sets of target links and attacking in an alternating fashion.

Because the Crossfire attack aims to congest the target links with low-rate \emph{benign} traffic, neither signature based Intrusion Detection Systems (IDS) nor alternative traffic anomaly detection schemes are capable of detecting malicious behavior on individual flows. The Crossfire attack's detectibility can even be further reduced by integrating any of the following features into the attack: the attacker (a) gradually increases bot traffic intensity, (b) estimates the decoy servers' bandwidth to avoid exceeding their bandwidth, (c) evenly distributes the traffic over the decoy servers, (d) alternates the set of bots flooding a target link, and (e) alternates the set of target links \cite{6547106}.

Although these techniques further sophisticate the attack, research described in this paper focuses on the effort the adversary has to invest for successful attack preparation and execution. As it turns out, the inherent complexities of the attack create also substantial execution obstacles, which exposes the attack to detection vulnerabilities. 

%For instance, a successful attack highly depends on how the network is structured and how the attacker initiates the attack sequence. 

In this paper, we describe how the Crossfire attack has been replicated in a realistic test bed emulation. The traffic has been measured during the topology construction phase and attack phase and analyzed for patterns and vulnerabilities of the Crossfire attack. The results indicate that characteristic traffic anomalies emerge in the attack region. Particularly, we found a correlation between coordination of the botnet traffic and the quality of the attack and a correlation between the attack distribution and detectability of the attack. Additionally, we show that due to the bot synchronization there is a warm-up period after the attack is launched and before the target links are overwhelmed. Because of this warm-up period and  the distinguishing patterns in the topology construction phase, the obtained results pave the way for novel detection methods in the early stage of the attack, when the attack traffic is formed \cite{misra2017early}. 

As a consequence, based on intrinsic property of attack traffic distribution, we propose a new approach to monitor the traffic volume (or intensity) on specific network regions for any sudden subtle changes on some of the links. Depending on the resolution of the monitoring scheme, this leads to an early detection of the attack, which we illustrate in this paper.

This paper also provides a functional analysis on how to assess the impact of the Crossfire attack on the effected area more realistically instead of over-estimating resources needed for attack detection and mitigation. We analyze these challenges in attack preparation and execution of the Crossfire attack and exploit them for attack detection. Hence, we describe a prototypical Crossfire attack detector, which exploits these vulnerabilities. For this, we utilize two supervised machine learning approaches: Support Vector Machine (SVM) and Random Forest (RF) for classification of network traffic to normal and abnormal traffic, i.e, attack traffic. To show the feasibility of detection, we report on the trained scenarios using the link volume as the main feature set. Finally, results of the attack detector are reported along with some future directions to improve the detector.

\section{Monitoring and Detection Approach}

\subsection{Monitoring points}

Considering the described Crossfire attack execution sequence, it turns out that there are potentially four ways to detect the attack: (a) detection at the traffic flows origin, i.e., bot sides, (b) detection at the target area, (c) detection at the target link, and (d) detection at the decoy servers. Following, we address the advantages and disadvantages each of the four ways to finally justify our choice for traffic monitoring. 

\begin{itemize}
\item \textit{Detecting at origin} can be the fastest way to stop an attack before even it is initiated. However, versatility and spatial distribution of bots (source of the attack traffic) makes it the most challenging option.%Attacker always can find a different set of bots from any region in the world.
\item \textit{Detection at target area} is the most reasonable approach as any target areas should be equipped for self defense. However, assuming not all decoy servers are inside the target area %but in the surrounding area
, early detection is impossible \cite{TLF}. %Responding to the attack when the target area is already isolated is hardly possible. \cite{TLF} provides an algorithm to localize the target link from target area after the area is disconnected.
\item \textit{Detection at target link} might be the simplest form of detection as simple a threshold based detection system that could detect the trend of the incoming traffic. %However, at the time that such algorithm can detect the overwhelming arriving traffic, it might be too late for reaction. Another important challenge on detection at target link is the lack of collaboration among ISPs. To cut down a target area, a set of links needed to be congested (not only one single link). %Most likely, each link in the set of the target links belongs to different ISPs. To truly detect a Crossfire attack, those ISPs need to share data and collaborate together. ISP collaboration is a hard problem to overcome.
\item \textit{Detection at decoy servers} can be the best approach to detect Crossfire attack. Assuming the target area is not far from the decoy servers (3 to 4 hops \cite{6547106}) detecting at the decoy servers might reduce the impact of the attack. %severeness of the attack. %Although it is claimed in \cite{6547106} that detection at a decoy server is hard, we believe it is not if a collection of decoy servers are investigated instead of only one. Since the attack traffic is evenly distributed among decoy servers and it is important to keep it this way \cite{6547106}, at certain time that the attack traffic reaches the decoy servers the correlation of the servers increases with the intensity which depends on the volume of the attack traffic and the arrival time of the attack traffic to each decoy server.% (see section~\ref{sec:early-detection}). Therefore, there are two main parameters which influences the detection which are volume of the attack traffic and the timing of the attack (attack coordination). We will investigate the effect of these two parameters on the quality of the detector.
\end{itemize}

%MERGE FROM OTHER PAPER: (I would start with: To summarize above points: Recall that the Crossfire attack is a distributed attack at both origin and destination sides, monitoring the traffic volume (or intensity) on several links of the network is advised. The detector looks for as specific global sudden characteristic change on some of these links. Intuitively, collecting more information from the network (hoping to have greater overlap with the attack network) increases the accuracy of the detection. However, to have an effective attack, the decoy-servers should be relatively close to the target area. Therefore, most probably a significant part of an attack traffic could be found in a closed area. Assuming an ISP is the major provider of the Internet of that area, we make our focus on a smaller scale of the monitored data from a single ISP. In Section\ref{sec:detection}, we will show that even with fewer number of monitored links, the accuracy of the detector is in acceptable range.

Our approach is based on detection at the decoy servers, because it is the exclusive area that the defender can detect the attack while actively respond to it. To emphasize the effectiveness of our detection approach at the decoy servers, we address the question of where is the best location to probe the network. In a high resolution, this probing can be placed either at the target link, before target link or after the target link. Monitoring a single link as a target link is not considered as a solution because of two reasons:
%MERGE FROM OTHER PAPER: Thus, assuming single ISP link probing\footnote{Multi ISP link probing requires collaboration among ISPs which is out of the scope of this paper}, in a very high resolution, the information could be gathered either from an ISP before or after the target link. Monitoring a single link (focus on the target link) is not in our interest because:

\begin{itemize}
\item Any links can be targeted for an attack. Therefore, there should be one-to-one detector for every link in the network. While, in our proposal there is only one detector but many probing points.
\item Monitoring and detecting based on a single link will fail in distinguishing between link attack and flash-crowd.
\end{itemize}

%OLD: \begin{itemize}
%\item Any links can be targeted for an attack. We as defenders do not have any 
%prior knowledge of the structure of the attack (at least this is the assumption).
%Therefore, to make sure that all attacks can be detected, there should be
%a one-to-one detector for every link in the network. 
%\item Detection based on monitoring a single link will fail in distinguishing
%between link attack and flash-crowd (non-malicious link failure). 
%\end{itemize}

The main goal is to detect the Crossfire attack without necessity of having the target link info. To find out the best monitoring domain, we assume to know the location of the target link for now. The question is which side of the target link provides more information for detection?
Assuming the number of ports of a switch/router is limited, considering only the immediate links before or after the target link might not help to choose a side. However, getting farther away from the target link the distribution of the intensity of the traffic on the links might be a function of the distribution of the end points. We will show in Section \ref{sec:dist} that more distributed attack traffic is more difficult to detect.

%Before addressing this question, it is worth to mention that as the Crossfire attack is a link flooding attack, a nice feature to work on for detection is the link utilization. Now, it can be argued that the link utilization of the links at the decoy server side might be under influence of the attack more than the links at the source of the attack. However, getting farther away from the target link the distribution of the intensity of the traffic on the links might be a function of the distribution of the end points.

Depending on the budget of the adversary, the number of bots purchased for an attack can be in range of thousands to millions. If the source of the attack traffic, i.e., bots, is geographically spread out, the variation of the traffic volume on most of the links is very small (for many routes there might be only one or few attack flows before they are aggregated at the target link). That leaves only few link closer to a target link worth to examine. However, the chosen decoy servers should not be very far away from the target area (if they are not inside the target area). Since there are smaller number of destinations for the attack traffic than the number of sources of generating them, it can be assumed that the variation of the volume of the traffic caused by the attack traffic on the links after the target link is higher than the links before the target link. % (even at the edge of the network). 
Therefore, we suggest monitoring links around servers or data centers results in better detection than around clients. 

The approach of evenly distributing the traffic for decoy servers
\cite{6547106}, might even support the above reasoning and rather
make it simpler to detect some variation on the traffic volume on
several links. The important element in this method is to be able
to monitor the traffic at several links and send the information to
a detector for decision making.

%Nevertheless, this hypothesis could fail if the decoy servers are selected from a highly spread area. Therefore, we need to examine the effect of the distribution of the decoy servers on the detectability of any jump in the traffic volume.
% The method we have discussed here will be supported by experiments in the subsequent sections of this paper.

\begin{figure}
\centering \includegraphics[width=0.90\linewidth]{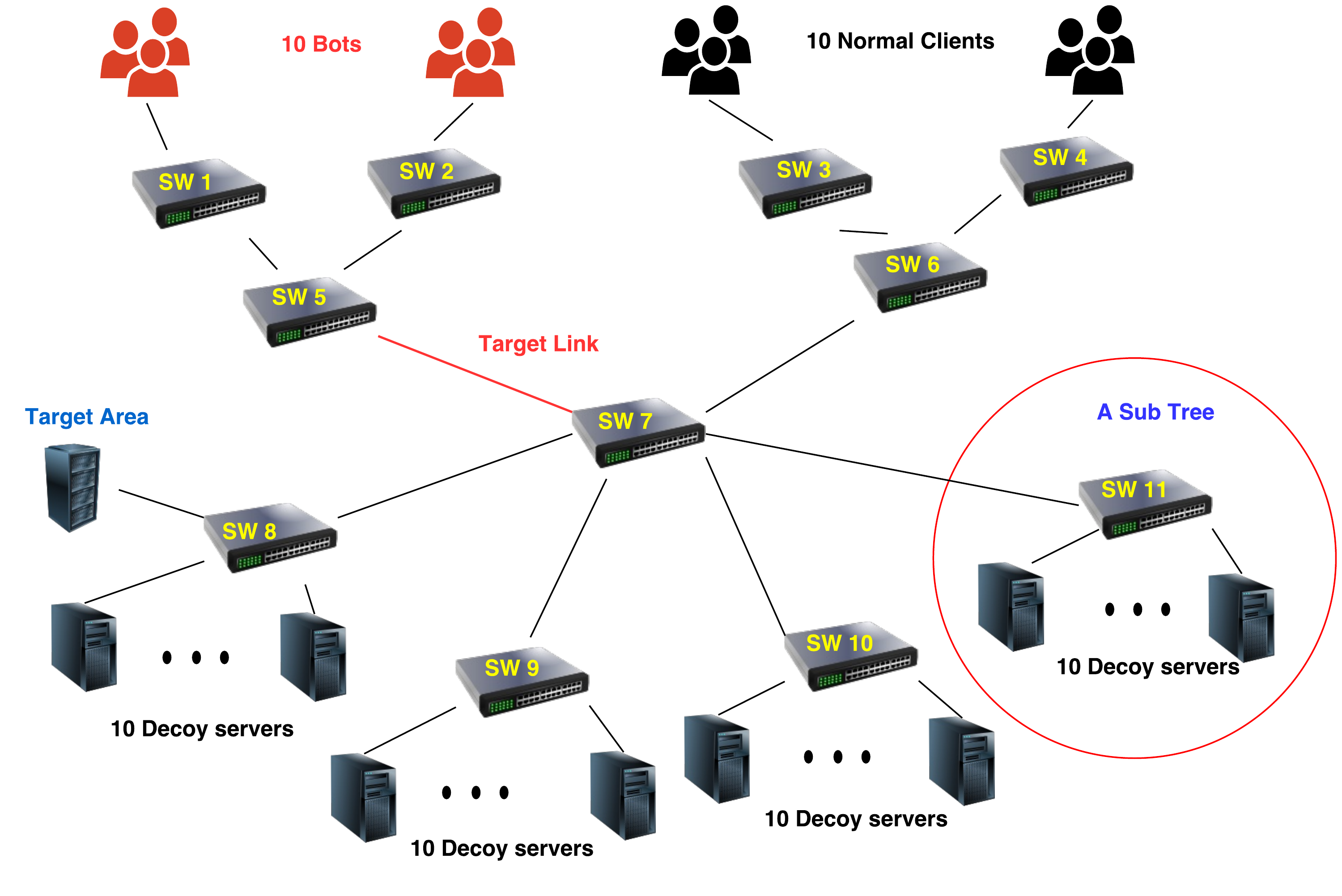}
\caption{\label{fig:topology}A four sub-tree topology of 10 bots, 10 normal clients, and 40 decoy servers. Each 10 decoy servers connected to a switch is called a sub-tree.}
\end{figure}

\section{Emulation Setup}
\label{sec:emulation_setup}

In order to substantiate our discussion from Section~\ref{sec:em_result},
we emulate the Crossfire attack in a realistic test bed environment. The test bed is implemented in Mininet and the following setup has been chosen for the emulation environment:

\begin{itemize}
\item SDN network created in Mininet.
\item SDN switches with POX controller.
\item D-ITG traffic generator \cite{DBLP:journals/cn/BottaDP12}.
\item Tree topology (cf. Fig.~\ref{fig:topology}).
\item Link bandwidth is set to 2 Mbps with 10 ms delay. 
\item POX controller gets link status every 5 sec from switches.
\item Bots generate both normal and bot traffic.
\item Some background traffic from clients. 
\item Some background traffic at the leaf switches (to decoy servers) to
level up the traffic at the edge links.
\end{itemize}

One focus in this paper is the correlation of the traffic distribution on the detectability of the Crossfire attack. We hence used a tree structure as the topology of the network. This permits us to intuitively expand the topology of the network, i.e., the tree structure, in order to widen the traffic distribution. Fig.~\ref{fig:topology} illustrates a base network for our emulation in which there exists several sub-trees, each of them includes $10$ decoy servers. To investigate different traffic distributions on the network, we design three variations of this topology as shown in Table~\ref{tbl::variation-of-topology}. Fig.~\ref{fig:topology} depicts the network topology of \textbf{4ST} which includes $4$ sub-trees, $11$ switches, and $40$ decoy servers. 

\begin{table}
\centering
\caption{Different variations of the Network Topology.\label{tbl::variation-of-topology}}{
\begin{tabular}{|c|c|c|c|}
\hline 
 & \# of sub-trees & \# of switches & \# of decoy servers\tabularnewline
\hline 
\hline 
\textbf{2ST} & $2$ & $9$ & $20$\tabularnewline
\hline 
\textbf{4ST} & $4$ & $11$ & $40$\tabularnewline
\hline 
\textbf{8ST} & $8$ & $15$ & $80$\tabularnewline
\hline 
\end{tabular}}
\vspace{-1.2em}
\end{table}

From practical aspect, Mininet with D-ITG traffic generator have limitation on the size of the network in the emulation. This is attributes to the fact that we need to reduce CPU utilization in SDN networks. Hence, the bandwidth of all links in our emulations are set to $2$ Mbps to be able to saturate the target link with fewer bots and less number of traffic generators. Moreover, the number of clients and bots are set to a small number of 
10 each, to compromise for a larger number of decoy server. Nevertheless, bots and clients can generate traffic in higher rate to rectify the problem.

All switches are SDN switches connected to a POX controller.
We modified the POX module \textit{flow\_stats.py} provided in Github~\cite{portstats}, which gives the controller its ability to collect some port- and flow-based statistics from switches. By using this code, the controller sends a stat request to all of the switches connected to the controller every five seconds. The respond from switches is the number of packets in the buffer of each port and the number of flows at each link. %(the detailed information of the flows are provided as well). 
There are 20 clients in this network including 10 bots (connected
to switches 1 and 2) and 10 normal clients (connected to switches
3 and 4). Clients can be considered as super clients which can 
generate traffic with higher rate than a normal client (or bot) can
do. %Since there are limited resources in the emulation environment and the focus of the study is on the distribution of decoy servers, we had to limit the number of clients in our emulation to possibly increase the number of decoy server.

Bots generate two types of traffic: normal traffic from beginning to the end of the experiments, and bot traffic which starts after {\it d} seconds and for duration of another {\it d} seconds. For experiments in Section \ref{sec:detection}, \textit{d} is set to 5 and to 30 minutes to have enough samples for the detector. 

There is a limited number of traffic types for both normal and bot traffic. Table~\ref{tbl:traffic} presents all traffic types used in the experiments.
Background traffic (normal traffic) consists of five application 
traffic including: Telnet, DNS, CSa (Counter Strike active player), VoIP and Quakes3, that D-ITG allows us to generate~\cite{DBLP:journals/cn/BottaDP12}. Both normal clients 
and bots are using these traffic to generate background traffic.
However, since we could not specify any inter-departure time nor 
packet size using these applications, we use simple TCP requests to 
generate attack traffic. To make the two type of traffic indistinguishable,
we add the same TCP traffic to the set of background traffic as well. This is the requirement of the Crossfire attack in which background traffic is indistinguishable from the attack traffic.
%Normal users generate 2 types of normal traffics. Although
%bot traffic generates attack traffic to saturate the target link,
%in nature there is not much different between normal traffic and bot
%traffic. Therefore, the size of the packets we use for normal and
%bot traffics are in the same range. 

Although the type of the normal and abnormal traffic should be the same,
the rate of traffic for the two type of traffic can be different.
In reality, the rate of bots' traffic must be
engineered by the attacker. Here we set the rate base on the remaining
bandwidth of the targeted link after receiving the normal traffic.
%However, to have a variety of traffic volume at each link, the size of the packet and the packet generation rate for each traffic type is a random number. 
%The range of the random number for packet size is the same for both normal and bot traffics. However, the range of the packet generation for the two types of traffic are different. Rate of packet generation for bots depends on the number of decoy servers in the experiment.
%(Observe that in the Crossfire attack the bot traffic is considered
%to be similar to the normal traffic.) To justify this, the bot traffic
%we are generating is the aggregated of several bot traffics at this
%point. D-ITG can generate uniformly distributed packet size and traffic
%rate by setting their maximum and minimum range. Depending on the
%size of the network (2 sub-tree or more), the rate of the packet generation
%varies. 
The details of the traffic types and their parameters for the setup in Fig.~\ref{fig:topology} is given in Table \ref{tbl:traffic}.

\begin{table}
\centering \caption{Parameters of the traffic.}
\label{tbl:traffic} 
\begin{tabular}{lcccccc}
\hline 
 & \multicolumn{2}{c}{pkt size (Bytes)} & \multicolumn{2}{c}{rate (pkt/sec)} &  & \tabularnewline
\hline 
Type  & min  & max  & min  & max  & Protocol  & Duration\tabularnewline
\hline 
Telnet & - & - & - & - & TCP & 60min \tabularnewline
DNS & - & - & - & - & TCP/UDP & 60min \tabularnewline
CSa & - & - & - & - & UDP & 60min \tabularnewline
VoIP & - & - & - & - & UDP & 60min \tabularnewline
Quake3 & - & - & - & - & UDP & 60min \tabularnewline
Bot  & 100  & 2000  & 15  & 200  & TCP  & 5min \tabularnewline
Background1  & 50  & 1000  & 1  & 80  & TCP  & 60min \tabularnewline
Background2  & 10  & 2500  & 1  & 80  & TCP  & 60min \tabularnewline
%Extra  & 5  & 5000  & 10  & 100  & TCP  & 15min \tabularnewline
\hline 
\end{tabular}
\end{table}

In addition to the traffic generated by the clients and bots, there
are extra traffic generators attached to some of the switches (mostly leaf switches) to increase the level of the background traffic at the links. This can be considered as the traffic coming from another part of the network which is not in Fig. \ref{fig:topology}. Since, the number of clients is much less than the number of servers, the extra traffic generators help to boost the level of traffic at the edge links connected to the servers.

\section{Emulation results}
\label{sec:em_result}

As noted before, the contribution of this paper is to expose hardships of the Crossfire attack and use them for an early detection method. We specifically focus on the effect of bot traffic synchronization on the quality of the Crossfire attack, and the effect of the distribution of the attack on detectability of the attack, which we describe in the following two subsections.
%\begin{figure}[t!]
%\centering \includegraphics[width=0.45\textwidth]{Images/extended_topo}
%\caption{\label{fig:ext_topology}The expanded network from 2 leaf branches
%to maximum 8 branches.}
%\end{figure}
Since our focus is on the detection of the attack, we ignore the first few steps of the Crossfire attack such link map construction, finding link persistence, or target link selection. We assume that all attack preparations have been made and the attacker is ready to attack. % (as explained in Sec. \ref{sec:emulation_setup}).

\subsection{Bot traffic synchronization}

% \begin{figure}
% \centering 
% \includegraphics[width=0.5\textwidth]{Images/targetLink}
% \caption{\label{fig:sync}The effect of bot-traffic synchronization on the warm-up period.}
% \end{figure}

The topology we use is presented in Fig. \ref{fig:topology}. At this stage, to bring down the target link, the adversary only needs to start the bot traffic and direct it to the decoy servers. Thus, the bot-master initiates the attack by sending the attack order to the Command and Control (C\&C) server or some selected peers depending on the structure of the botnet. Bots usually update each other in a polling or pushing mechanism. However, the question which is of interest is what happens if bots receive the attack order in different time order?

When designing Crossfire detection mechanisms, an often ignored part of the Crossfire attack is the phase from the attack initiation and the successful impact of the attack \cite{6547106}. This often ignored part of the Crossfire attack, which we call it \textit{warm-up period}, is the time difference between the time of the first bot-flow of the attack reaches the target link and the moment the target link is down. By definition, the attack actually happens at the end of the warm-up period when the target links are down. Since, reaching a zero time warm-up period is hard, this period can be used for early detection and before the attack successfully takes place.

In fact, for several reasons reaching a zero warm-up time is hard. One reason could be the dynamic delay of packet arrival at the target link. That could be because of variations of hop distances from bots to target link, or the delay in receiving attack order from the adversary. Any sudden significant change on traffic volume can be detected by firewalls and IDSs. Therefore, adversaries gradually increase the attack traffic   volume to prevent being detected. To have a \emph{perfect} link failure, the volume of the traffic arriving at the target link should be slightly higher than the bandwidth of the target link itself. However, this might not happen immediately. There are three main reasons for gradual traffic growth: 

\begin{enumerate}
\item Bot traffic can be originated from any geographical location in the world and they might arrive at the target link with different delays
(dynamic delay). 
\item Since the source of the attack is a botnet, it is reasonable to assume that there is some time slack between each bot to start sending the bot traffic. This time slack can be caused by how bots receive updates from their C\&C center or from other peers in an advanced P2P botnet, but also from the malware itself~\cite{p2p,Wu_detectingpeer-to-peer}. 
\item Bots might gradually increase their traffic intensity to prevent detection~\cite{6547106}. This can be considered as the main reason of gradual increasing the attack traffic volume.
\end{enumerate}

\begin{figure}
\centering
\includegraphics[width=0.90\linewidth]{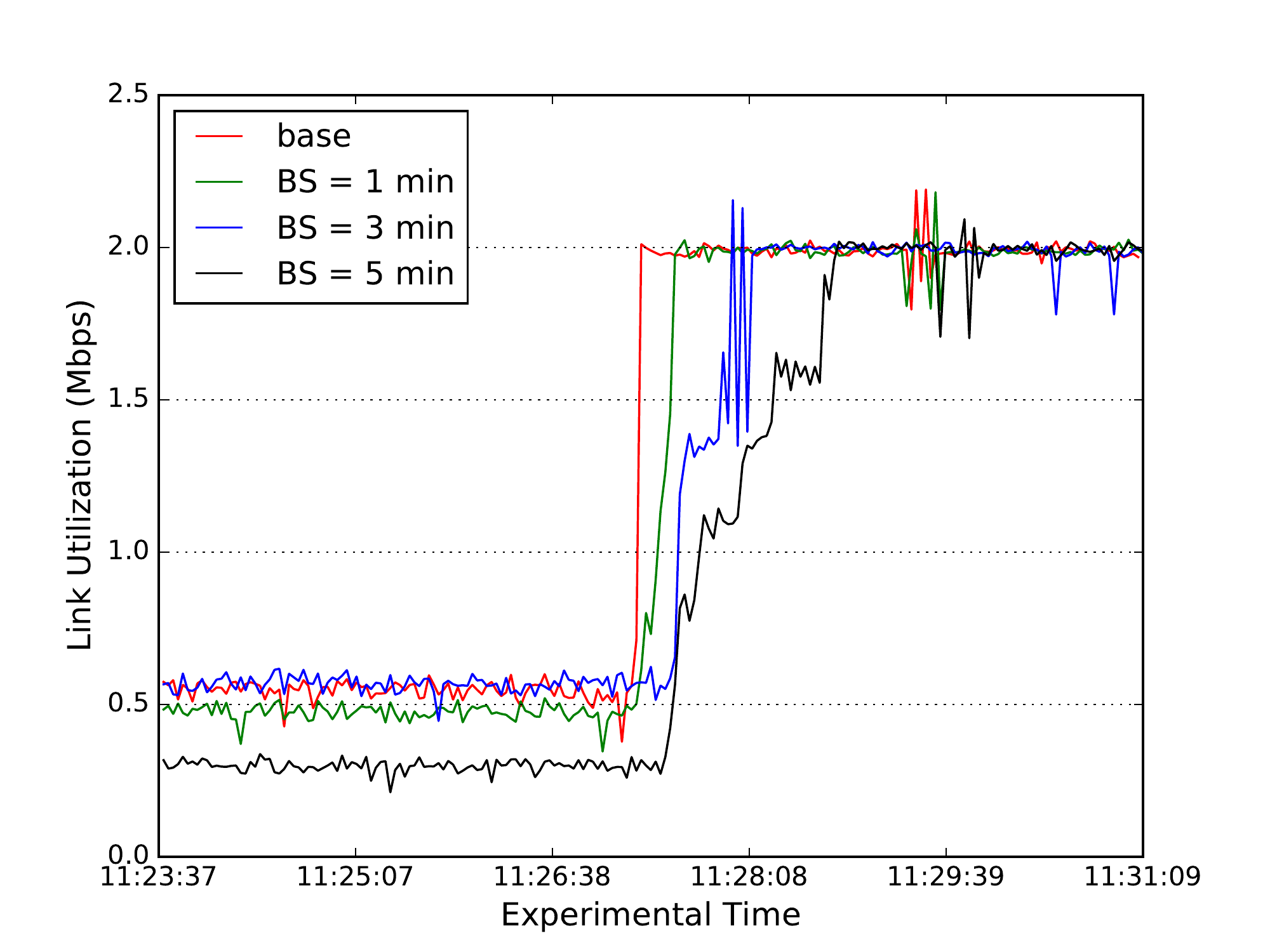}
  \caption{The effect of bot-traffic synchronization on the warm-up period. 2 Sub Trees with different warm-up periods.}
  \label{fig:sync}
\end{figure}

To illustrate the effect of the bot synchronization on the traffic volume of the target link, the result of an emulated attack is presented in Fig.~\ref{fig:sync}. A two sub-tree version of Fig.~\ref{fig:topology} is used to generate above results. At this stage, to bring down the target link, the adversary only needs to start the bot traffic to the decoy servers. That means, at this stage, we are only running the last part of the Crossfire attack. 
%Thus, the botmaster initiates the attack by sending the order to the C\&C server or some selected peers (In a peer-to-peer botnet) depending on the structure of the Botnet.
%However, the question which is of interest is what happens if bots receive the attack order in different time order? Or what is the effect of the dynamic delay (the delay caused by the distance of the bot to the target link) on the attack flow?
%Alternative: To understand the effect of inexact bot synchronization, bots start to send their attack traffic in a time margin which is called Bot Starting (\textit{BS}) attack time. To begin the attack traffic, each bot takes a random number between 0 and \textit{BS} as the starting time of generating attack traffic.
Fig.~\ref{fig:sync} illustrates the utilization of the target link before and after the attack. Different curves in different colors represent different \textit{BS} time for bots to generate the attack flow. 
%The blue curve is the baseline to show that perfect attack happens when all bot traffic simultaneously arrive at the target link with their maximum intensity. In other curves, bots pick a random number in range of mentioned \textit{BS}  value to start sending the attack traffic.

In Fig.~\ref{fig:sync} the red curve is the baseline to show that perfect attack happens when all the bot traffic simultaneously arrive at the target link with their maximum intensity. The time interval which is used in \textit{BS} in above experiments is in range of 1 to 5 minutes. The reason is that in a p2p platform (the most recent platform to synchronize Botnets) peers usually contact each other in range of few minutes~\cite{p2p,Wu_detectingpeer-to-peer}. For instance, Skype peers update only closer peers every 60 seconds~\cite{p2p}. In other studies like~\cite{cicadas}, the time synchronization between bots is reported in range of few \textit{milliseconds}. However, there are few steps (three state machine) before they can reach to that accuracy and those states take sufficiently long (i.e., few minutes). Therefore we still can assume that there is enough time in range of few minutes before the real attack takes place. %It is worth to emphasize that what 
We are looking here at the time difference between arrival of the first packet of each bot to the target link. The time difference between arrival of each packet of any bot traffic could be in range of milliseconds which is not our concern here.

Since we are now aware of this possible early detection, we discuss how to detect an attack which is formed by some low-intensity \emph{non-malicious} traffic. The main idea is analogous to detecting the variation of the volume of the traffic at several links. Since one cannot gain any information from per flow traffic monitoring (the attack traffic is benign traffic), and attack type is a flooding attack, probing the volume of the traffic at several links might be effective. Although, a single bot-flow is very small and can be detected neither at IDS nor at the server, the aggregation of the flows are not small anymore. All these small traffic flows must be aggregated at the certain time and place to be able to overwhelm the target link(s). 

\begin{figure}
\centering \includegraphics[width=0.97\linewidth]{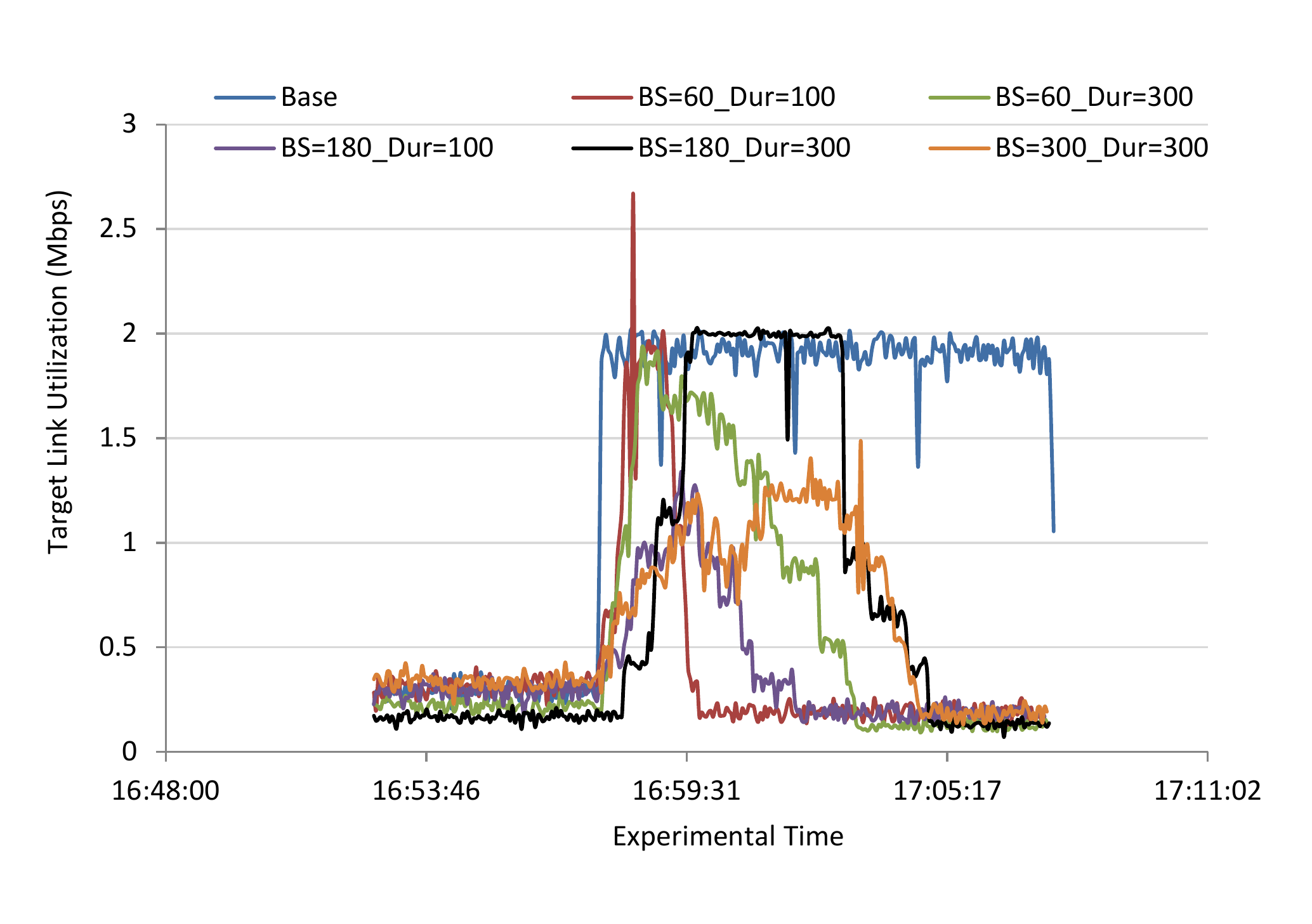}
\caption{\label{fig:sync_dur}Bot traffic with various starting points and traffic flow duration.}
\end{figure}

Another important parameter to generate the bot traffic is the duration of the attack. Usually, bot-masters (adversaries) tend to reduce the duration of the attack to prevent being detected. The combination of the dynamic delay and the attack duration is difficult to figure out. The attack duration parameter which is the time difference between the end of the warm-up period to the end of the attack, is named \textit{Dur} in our experiments.

In the case of the Crossfire attack, a rolling mechanism is introduced to keep the attack at the data plane (evade activating control plane which redirects the traffic) \cite{6547106}. In the rolling scheme, a set of target links only be used for a specific period of time before it switches to another set of target links. The duration they used in the rolling scheme is \textit{3 minutes}. The 3 min is the \textit{keep-alive} messages time interval for the BGP algorithm. This duration might be insufficient when the attacking traffic is gradual because of any of the above mentioning reasons.
%due to poor bot synchronization. 
%Poor bot synchronization
This limitation of small duration of the attack forces the adversary to introduce another delay in forming the attack which causes a larger warm-up period. The effect of the length of the attack duration (including warm-up period) with various \textit{BS} and \textit{Dur} parameters are depicted in Fig.~\ref{fig:sync_dur}. For comparison purpose, the baseline is the case where bots simultaneously generate traffic (warm-up period is zero) for an unlimited duration of time. For the other curves, there are warm-up periods for \textit{BS} length and duration of length \textit{Dur}. 
%Like \textit{BS}, \textit{Dur} is a
%uniformly generated random number with a specific minimum and maximum
%values. The minimum value is fixed to \textit{60 sec} and the maximum
%value is the number mentioned for each curve in Fig. \ref{fig:sync_dur},
%as \textit{Dur}. We use a random start and end time for each bot traffic
%to make it even harder to detect. 

Figure~\ref{fig:sync_dur} shows that, with less synchronized attack traffic,to have a successful attack, either the duration of the attack should be prolonged enough to pass the warm-up period or, the adversary should delay the attack and let the warm-up period passes before initiating the attack. 

Although, the parameters in our experiments are set to small numbers (few minutes of warm-up periods)\footnote{There are some technical settings that can be used to support the selection of the small parameters, such as, in a p2p platform (the most recent platform to synchronize botnets) peers usually contact each other in range of few minutes \cite{p2p,Wu_detectingpeer-to-peer}, or Skype peers update only closer peers every 60 seconds~\cite{p2p}.}, the result can generally be extended for longer periods. The main reason of keeping the simulation time short is limitation of resources in our setup. Increasing the size and the time of the experiments reduces the accuracy of the traffic generator~\cite{DBLP:journals/cn/BottaDP12}.

\begin{figure}
  \centering
  \includegraphics[width=0.90\linewidth]{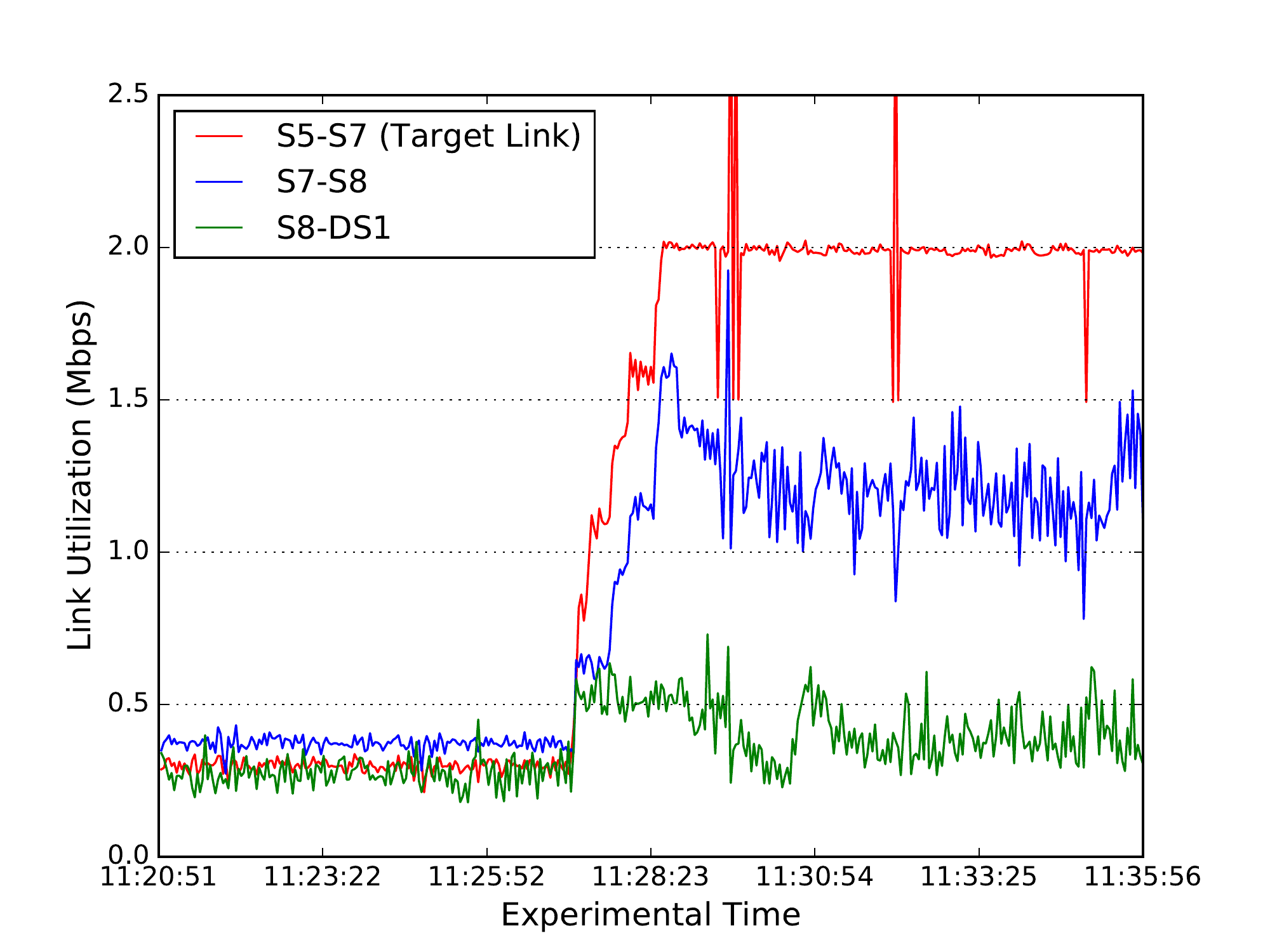}
  \caption{2 Sub-Trees where jump can be seen at all levels.}
  \label{fig:2sub}
\end{figure}

\subsection{Distribution}
\label{sec:dist}

We discussed the traffic synchronization problem in forming 
the attack. The introduction of warm-up period can be used for early detection of the Crossfire attack which is the topic of the next section. In this section, we introduce a hypothesis about the link traffic intensity variation caused by the Crossfire attack and suggest to use it for detection. We hypothesize that even if the Crossfire attack is successfully formed by generation of very low intensity attack traffic, unavoidably there will be a sudden jump in the traffic on (backbone) links, whereby this jump will be characteristic for a Crossfire attack. 

%\textit{\textbf{ Hypothesis:}  Even if the Crossfire attack successfully generates very low intensity attack traffic, unavoidably there will be a sudden jump in the traffic on (backbone) links, whereby this jump will be characteristic for a Crossfire attack.}

\begin{figure}
    \centering
 %    \begin{minipage}{0.5\textwidth}
%  \centering
  \includegraphics[width=0.90\linewidth]{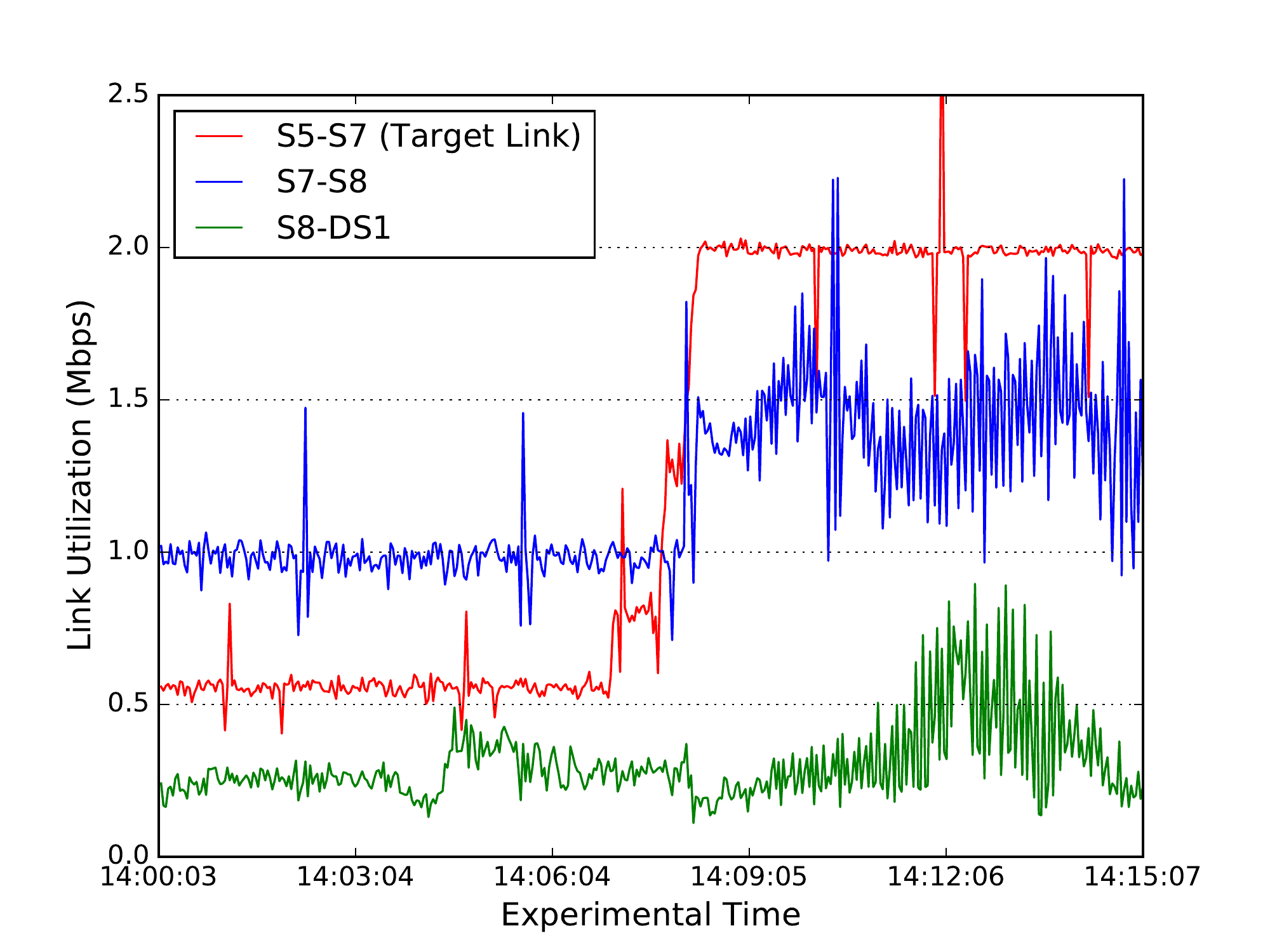}
  \caption{4 Sub-Trees and jump still visible.}
  \label{fig:4sub}
\end{figure}%
%    \caption{Switch 7 is connected to multiple sub-trees. In (a) the jump can be seen at all levels. Increasing the number of sub-trees to 4 subs in (b) reduces the jumps. with 8 sub-trees in (c) the jump at the last link is harder to recognize.}%
The main objective of the Crossfire attack is to bring down a set of target links to effect the connectivity of a target area. Depending on the power of the attacker, the target area could be cut off completely from the Internet or the quality of the connection to the Internet could be degraded. Either way, to bring down the target link the link utilization should be increased to its maximum capacity. The extra unwanted traffic at the target link must go through downstream links and affects their utilization. For instance, the attack traffic at the target link between switch 5 and switch 7 in Fig.~\ref{fig:topology} must pass through the four subsequent links between switch 7 and downstream switches 8, 9, 10 or 11. This sudden extra change on traffic has a huge impact on these downstream links and perhaps the effect goes down further to other links as well. We will examine this impact through emulation and report all the results for most of the links below the target link. Then by expanding the size of the network, we try to hide this impact by distributing the jump on the target link through more links.
%Due to some technical problems of the traffic generator, 
In these experiments, we do not consider the gradual traffic intensity increase at the bots. All bots send traffic at the maximum predefined level. The warm-up period is 3 min.

%\begin{figure}[t!]
%\centering \includegraphics[clip,width=0.5\textwidth]{Images/Dist_4}
%\caption{\label{fig:dist4}Switch 7 is connected to 4 sub-networks. The last
%link is the link to the Decoy Server 1 (DS1).}
%\end{figure}

% \begin{figure}
% \centering \includegraphics[width=0.90\linewidth]{Images/Dist_8}
% \caption{\label{fig:dist8}Switch 7 is connected to 8 sub-networks. The last link is the link to Decoy Server 1 (DS1); it is harder to recognize the jump on this link. %Here there is only one link for each level of the network.
% }
% \end{figure}

%\begin{figure}[t!]
%\centering \includegraphics[clip,width=0.5\textwidth]{Images/Dist_S8S7}
%\caption{\label{fig:distS8S7}Comparing the link utilization for the link connecting
%Switch 7 to Switch 8 in three different network sizes.}
%\end{figure}

\begin{figure}
  \centering
  \includegraphics[width=0.90\linewidth]{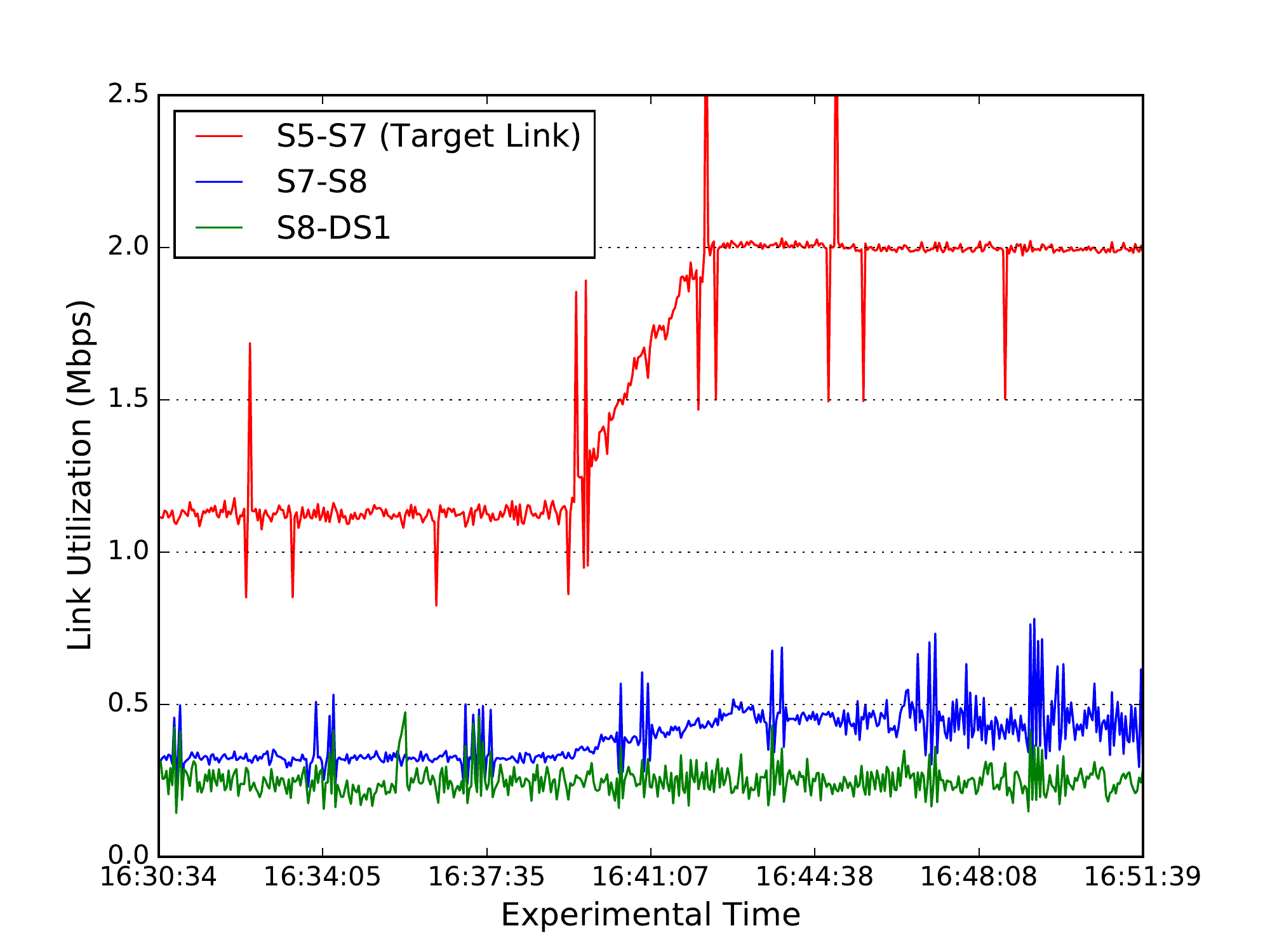}
  \caption{8 Sub-Trees where jump at the edge links are not visible.}
  \label{fig:8sub}
\end{figure}

The first scenario is a two sub-tree network of Fig. \ref{fig:topology}. In this scenario, the traffic at the target link can only go through two other paths. Fig. \ref{fig:2sub} shows that during the attack time, the target link is completely utilized and the other two links underneath of the target link are under influence of the sudden traffic change. The green line is for the traffic at the edge of the network where the switch 8 is connected to the Decoy Server 1. The result of running the similar experiments with some larger networks is illustrated in Fig.~\ref{fig:4sub} and Fig. \ref{fig:8sub}.

Comparing the link utilization in Fig.~\ref{fig:2sub}, Fig. \ref{fig:4sub} and Fig.~\ref{fig:8sub} shows that increasing the number of branches in the network, reduces the obvious jump on the downstream links. In particular in Fig. \ref{fig:8sub}, it is very difficult to distinguish the attack period only by looking at the edge link (the green line) utilization.

%Note that, in our experiments, increasing the number of branches increases the total number of decoy servers. That explains the scale of the traffic volume of normal traffic for the bigger networks. One conclusion could be that the attack is more efficient on larger networks. If a target link is already in the path of many traffic flows to many public servers, the smaller attack traffic is required to congest the target link. 
%Nevertheless, since the number of decoy servers connecting to each branching switches remains the same, there is not much difference on the normal traffic volume on the links after the target (congested) link in different network sizes. This can be seen in Fig. \ref{fig:distS8S7}.  
%Of course increasing the network size vertically instead of horizontally, the volume of the normal traffic of such links could be affected as well.

The main point of these experiments is to show that sensing a similar variation on traffic intensity on multiple links could be a good indication of Crossfire attack for detection. Since expanding the network reduces the jump in the link utilization, the detector must be accurate enough to detect very small variations of the traffic intensity where cannot be detected by unarmed human sight. %The next section discusses the detector along with some experimental results to show the parameters that influences its accuracy.

\section{Early Detection Method\label{sec:Early-Detection-Method}}

\subsection{Warm-up phase}\label{sec:early-detection}
\begin{figure}[t!]
\centering \includegraphics[width=0.90\linewidth]{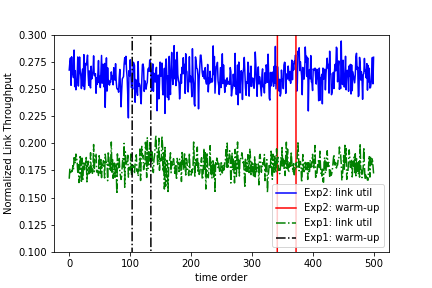}
\caption{\label{fig:visualCorrelation}Link utilization of one link with different attack intensity.}
\end{figure}

\begin{figure}[t!]
\centering \includegraphics[width=0.90\linewidth]{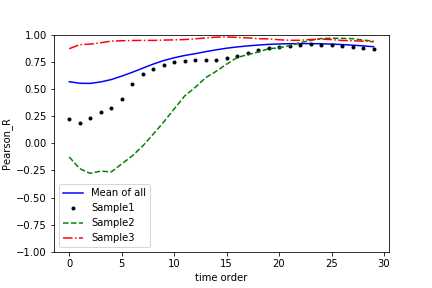}
\caption{\label{fig:corl_attack}Correlation among links when there are attacks.}
\end{figure}

As mentioned, warm-up period is the phase from the attack initiation and the successful impact of the attack \cite{6547106}. Early detection means detecting the attack during this phase when the attack traffic has reached to the decoy servers but the network is still operational. We show that by the time the attack starts the correlation among links to decoy servers gradually increases during the warm-up period potentially providing sufficient time and data to detect the attack.
%Thus, early detection means that detecting the attack during
%the warm-up period and before the attack successfully takes place.

For an effective and early detection, we propose to monitor the traffic
volume and intensity on several links of the network for simultaneously
occurring sudden characteristic change on some of these links. Based
on the awareness of this possible early detection, we discuss how
to detect an attack which is formed by low-intensity \emph{non-malicious}
traffic.

\begin{figure}[t!]
\centering \includegraphics[width=0.90\linewidth]{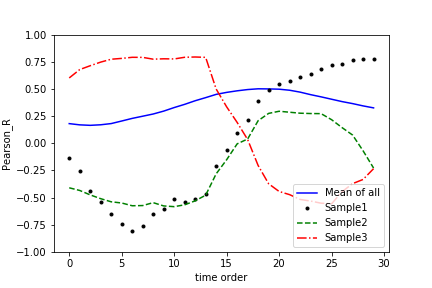}
\caption{\label{fig:corl_smallAttack}Correlation among links when the attack intensity is reduced.}
\end{figure}

\begin{figure}[t!]
\centering \includegraphics[width=0.90\linewidth]{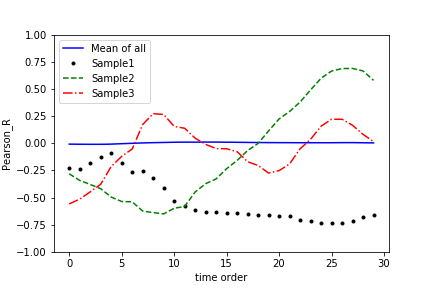}
\caption{\label{fig:no_attack}Correlation among links when there is no attack.}
\end{figure}
The main idea is analogous to detecting the variation of the volume
of the traffic at several links. Since one cannot gain any information
from per flow traffic monitoring (the attack traffic is benign traffic),
and attack type is a flooding attack, probing the volume of the traffic
at several links might turn out to be effective. Although a single
bot-flow is very small and can be detected neither at IDS (Intrusion Detection System) nor the server, the aggregation of the flows is not small anymore. All of these small traffic flows must be aggregated at the certain time and place to be able to overwhelm the target link(s). 
This variation on the traffic volume at several links correlates them more and this is where the attack can be detected.

% The expected jump could be lessened if the traffic is distributed trough more spread away decoy servers. However, there are few challenges for such a distribution: 
% \begin{itemize}
% \item For a more effective attack, decoy servers should be close to the target area (restriction on decoy server selection).
% \item Limited number of ports on a switch/router. 
% \item If the target link is distant from the target area, the attack will be rendered less effective.
% \end{itemize}
% Hence, one important question is how much freedom an adversary has in selecting the decoy server or the target link? If the attacker is forced to choose many servers from a single campus then the jump is inevitable.

\subsection{Experimental results}
%We believe it is possible to detect the Crossfire attack at the early stage of the formation of the attack because, the similarity of the attack traffic at the links to decoy servers, increases the correlation among the links which do not suppose to correlate. It is possible that two or even more links by a chance show similar behavior (load of the link at all of them increase or decrease) at a time, but it is not likely to observe this correlation among all of them.

%\textbf{explain how generate the results. a curve of pearson-r}
%To show that how the Crossfire attack correlates more the links under influence of the attack, the results of some specific experiments are reported here (figures \ref{fig:corl_attack} and 
%\ref{fig:corl_smallAttack}).
In this section, we present experimental results to support the hypothesis of early detection of Crossfire attack based on the correlation among the links to decoy servers.

The experiments in this section are different in a way that they are designed to study the correlation among the links with and without the attack. Thus, the attack traffic is not designed to overwhelm any target links. The objective of the attack traffic is to add extra scheduled traffic at all decoy servers.

\section{topology}
The same tree with 8-subtrees (80 decoy servers) and the same traffic types are used. In both experiments, there are a warm-up period of length 30 samples. 
%This is the duration that we are interested to detect the attack. 
The attack intensity during this period gradually increases at every time sample. This extra attack traffic for the first experiment (\textit{experiment-1}) increases from 300 bps to 600 bps and for the second experiment %(\textit{experiment-2})
increases from 60 bps to 150 bps.

The normalized (l1-norm)\footnote{l1-norm is used only for illustration purpose to preserve the the level of the link utilization at each experiment.} data of the link utilization of one link for both experiments is illustrated in Fig. \ref{fig:visualCorrelation}. The figure shows that the attack intensity for the first experiment (green curve) is higher than the second experiment (blue curve). Higher link utilization of the second experiment might hide the small variation of the attack traffic. The warm-up period for each experiment is highlighted with two parallel line.

Pearson-R is used to measure the correlation among the links.
Correlation is computed for every possible combination of two links. Since there are 80 decoy servers in our experiments, there are $\sum_{n=1}^{79} n = 3160$ combination of two links. Pearson-R returns a single value for two sets of data, representing how tightly (or loosely) the two sets are correlated together. However, we are interested in observing how correlation of two links for a duration of the warm-up period evolves. Therefore, Pearson-R is calculated for a window size of 30 (the same size of warm-up period) points. To calculate the first value of the Pearson-R, there are 29 sample points before the attack and one sample of the attack in the set. Then, the window is moved one sample to calculate the second value with 2 attack samples and 28 samples before the attack. Finally when the window reaches to the end of the warm-up period, all 30 samples in calculating Pearson-R include the attack traffic.

The result of experiment-1 is reported in Fig.~\ref{fig:corl_attack} 
%and for experiment-2 is reported in figure\ref{fig:corl_smallAttack}. 
This figure shows that the correlation constantly increases even for links that they are not correlated before the attack (sample-2, the green curve). 
The average of correlation among all links (the average over all 3160 pair of links) are presented and 
proves the positive effect of the attack traffic in increasing the correlation among the all links. 
%effectiveness of measuring the correlation.
Figure \ref{fig:corl_smallAttack} illustrates the result of experiment-2, 
When the attack intensity is reduced. This attack traffic is not strong enough to affect correlating on all links. For instance, the Pearson-R value of the two links in sample-3 curve of the Fig. \ref{fig:corl_smallAttack} are changing based on the background traffic on the links (they are not influenced by the attack traffic). However, there are some combination of links that are under influence of the attack traffic. Sample-1 curve in the same figure is one such example. 
We observed that the effectiveness of the attack traffic on the correlation of links is a function of the intensity of the background traffic. Smaller volume of attack traffic does not effect the correlation when there is a large amount of background traffic passing the link.
%In average, the correlation of all links are positive and more than 0.25 and less than half. 
%It should be emphasized that the attack traffic is not breaking down the target link in these experiments and 

The results of the link correlation when there is no attack traffic involved, is reported in Fig.\ref{fig:no_attack}. The figure shows that in average the correlation among the links are zero. Although, there might be some positive correlation among some links (like sample-2), this is not a general trend in the network.

%In overall, we believe there is a good chance for early detection of the Crossfire attack. The two characteristics of the attack which are, uniformly distributing the volume of the attack  traffic among all decoy servers, and the fact that the attack traffic must aggregate at the specific location and time, influences the correlation among network links.

%\textcolor{red}{The correlation is a function of the intensity of the attack traffic to the 
%background traffic level on the link. For instance in figure\ref{fig:corl_attack} although the volume of the attack is small and might be missed by an unarmed eye, still it is 
%effective enough to increase the correlation among all the links. 
%On the other hand, the attack volume chosen for simulation setup of 
%figure\ref{fig:corl_smallAttack} is even smaller which depending on the 
%original (background) volume of the traffic on the link, the extra traffic 
%might not have influence on the correlation between links or 
%it does. for some links like sample0 it increases the correlation among the two links, whilst it doesn't have effect on the 
%correlation between the two links in sample3.}

\section{Crossfire Detection with Machine Learning}
\label{sec:detection}

%1- motivation for using machine learning for attack detection (anomalous traffic behaviors)
%-2 Categorization of Anomaly and Normal

The Crossfire attack poses great challenges for security researchers and analysts both in detection and mitigation as the packets streaming from bots in the network are seemingly legitimate. While the objective of the Crossfire attack is to deplete the bandwidth of  specific network links, a distinct traffic flow between each bot to server, i.e., ``bot-to-server'' is usually very less intensive flow, %has very low flow 
and consumes a limited bandwidth at each link. Thus detecting a single flow (or very few number of them) at a link is hard to detect and filter.  On the defender’s side, Traffic Engineering (TE) is the network process that reacts to link-flooding events, regardless of their cause~\cite{liaskos2016novel}. As an attacker, we like to hide the variation of traffic bandwidth as much as possible from the TE module. 

\begin{figure}
    \centering
  \includegraphics[width=0.90\linewidth]{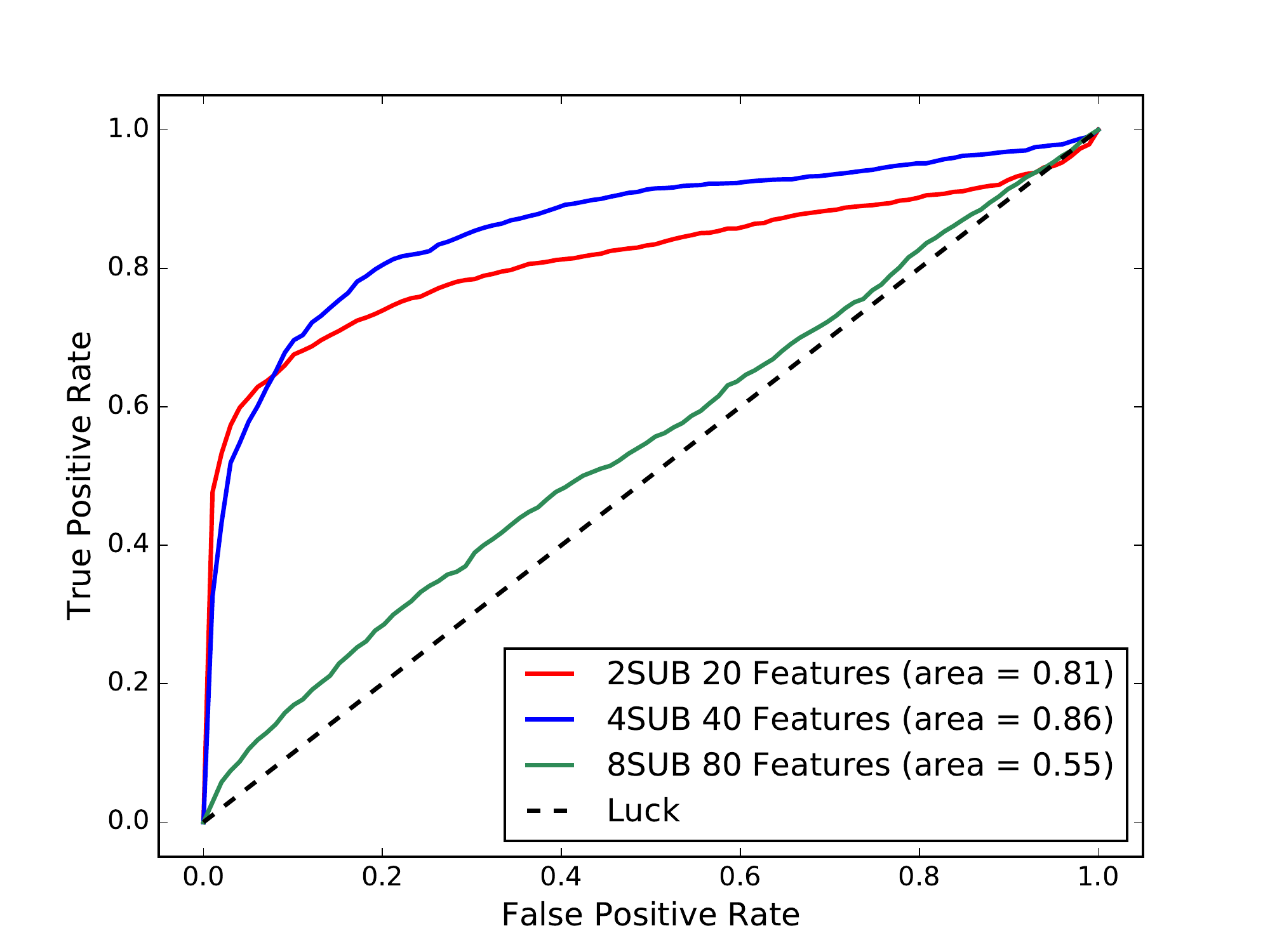}
  \caption{Classification result for SVM with different distribution.}
  \label{fig:allSubs}
\end{figure}%

The study in this section is to show that if the attacker distributes the benign traffic effectively enough, the defender face much trouble to distinguish the attack traffic from normal traffic when detecting the attack far away from the target link. 

Indeed, the analysis, in previous sections, have already attested the imminent importance of traffic distribution. Following this direction, we leverage state-of-the-art approaches in machine learning to investigate the effect of traffic distribution in concealing and detecting  the Crossfire attack from available traffic data. To do so, we utilize supervised learning for classification of network traffic to normal and abnormal traffic, i.e, attack traffic. 

\begin{figure}
  \centering
  \includegraphics[width=0.90\linewidth]{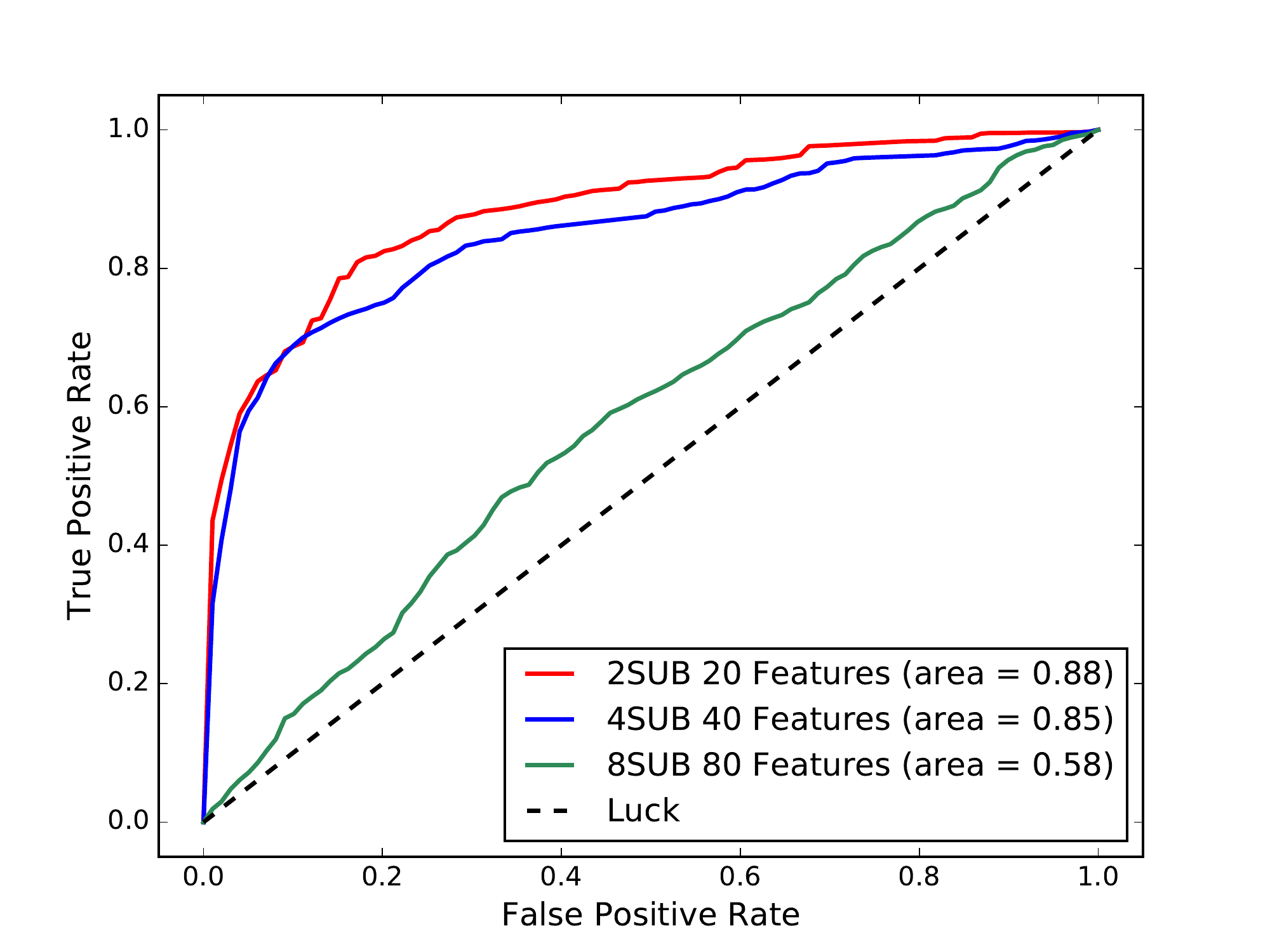}
  \caption{Classification Result for RF with different distribution.}
  \label{fig:allSubs:rf}
\end{figure}

%The intuition behind this study is that we attempt to construct a model from gig data collected from network. More specifically, we have utilized supervised machine learning to classify traffic to two classes of normal and attack data.

%Machine learning techniques are uses to classify malicious data and normal data [9]. Machine learning techniques learn from data that is gathered by statistical manner. This concept is come from data mining which collect data and discover the knowledge. Machine learning collect data from network. That collect behavior of nodes as per performance metric. 

%Machine learning learn from those data and classifies  different classes as linearly and known output in supervised learning. Main goal of this technique is to predict the future of those according the  behavior which learns from training set of data. 

\subsection{Learning models}
%-Two approaches including SVM and RF
%-Experimental Setting
%+ DataSet
%+Feature Extraction
%-Results
In this paper we attempt to construct a model from big data collected from network. We utilize two supervised learning approaches: Support Vector Machine (SVM) and Random Forest (RF) as they are commonly used machine learning approaches which have demonstrated effective performance on different datasets and problems. 

\paragraph*{SVM} 
It is known as one of the most powerful and non-probabilistic binary classifiers which attempts to separate the two classes of data with a hyperplane in a multidimensional space of features. We utilize linear-SVM due to its scalability.%~\footnote{We used the implementation of Scikit-learn}
%SVMs are non-probabilistic binary classifiers [28]. SVM is considered one of the most powerful supervised classification algorithm. It works by representing each feature vector in a multidimensional space and trying to find a linear separation (i.e., an hyperplane) for the classes. In some cases, however, a linear separation of the  space is not possible, hence it uses the so-called kernel trick, which implicitly increases the dimensionality of the space, resulting in an easier separation in a much higher dimensional space, due to the increased sparsity.
\paragraph*{RF}
Inspired by ensemble learning and bootstrapping, RF leverages multiple instances of decision trees, where each tree is built based on a randomly selected portion of training set. After computing the output of distinct trees, the final decision is made by aggregation of the outputs via a majority voting scheme. The Random Forest Algorithm was chosen because the problem of Crossfire detection has the requirements of high accuracy of prediction, ability to handle diverse bots, ability to handle data characterized by a very large number and diverse types of descriptors.%, ease of training, and computational efficiency. 

\subsection{Dataset and feature extraction}
We utilized an emulated dataset collected based on the experiments discussed in Section~\ref{sec:emulation_setup}. To generate the attack we used the topology designed and collected the data from distinct switches. As aforementioned, the objective of this section is to study the subtle variations in traffic data of the network to design effective detection approach for the Crossfire traffic. Therefore we employ the volume of traffic in different links of the network to construct feature vectors.  

We evaluate the performance of the learning approaches via the area under the receiver operating characteristic curve (AUC) ~\cite{powers2011evaluation}, which illustrates the true positive, i.e., sensitivity, as a function of false positive, i.e., fall-out. 

%Due to the imbalance nature of the problem, AUC provides a good explanation of the effectiveness of the proposed method.

\subsection{Experimental Results}
In this section, we design and analyze experiments to answer the following questions:
\begin{enumerate}
\item What is the impact of distribution of bot-to-server traffic in the performance of classification algorithms?
\item What is the impact of extracted features in the performance of classification algorithm?
\item What is the impact of levels of the links (in a tree structure) used for feature extraction?
\end{enumerate}

\begin{figure}
    \centering
 %\captionsetup{justification=centering,margin=0.5cm}
    \includegraphics[width=0.90\linewidth]{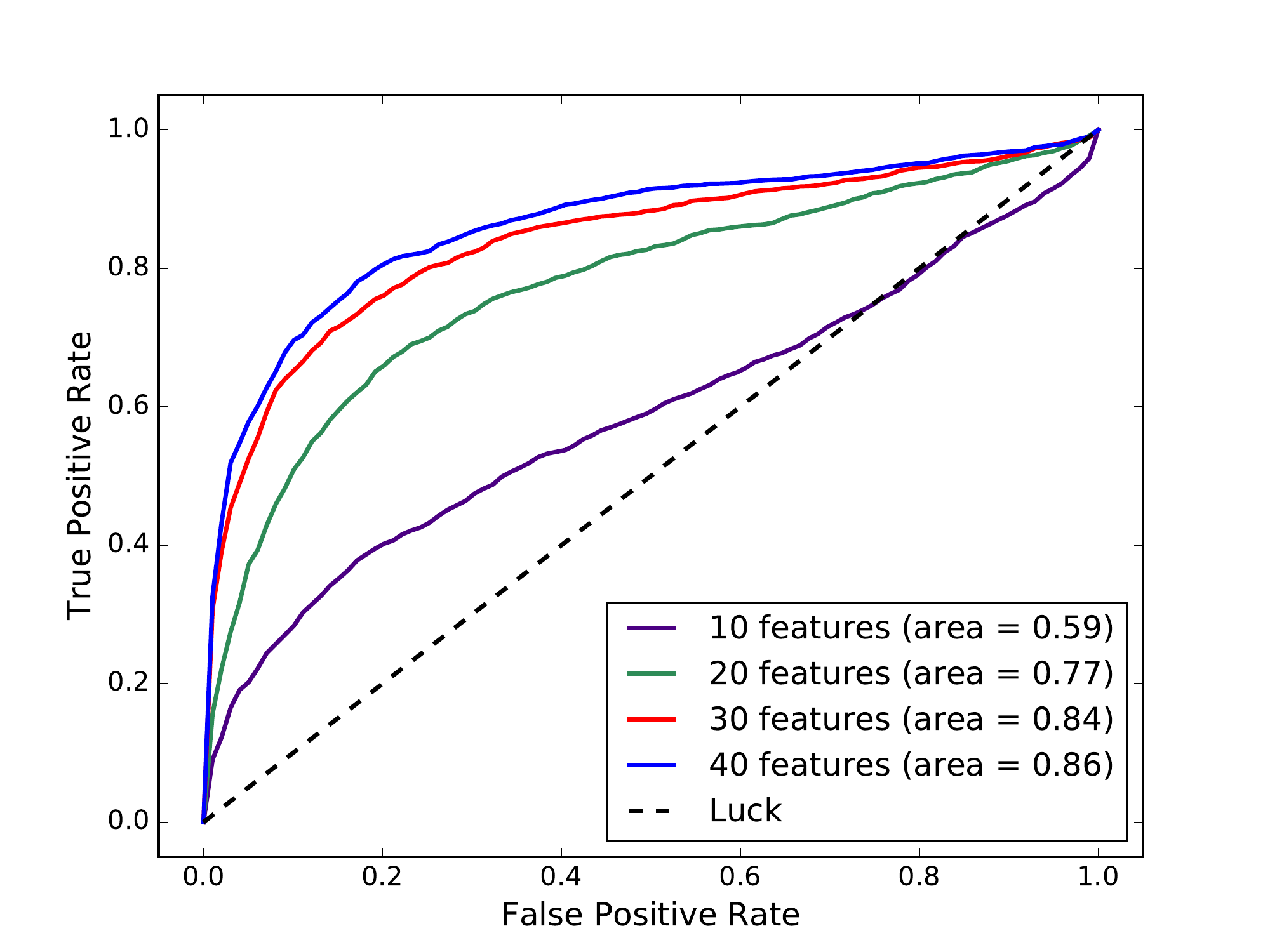}
  \caption{The effect of number of features on attack detection for 4-sub-tree.}
  \label{fig:4subFeatures}
\end{figure}

\subsubsection{The effect of traffic distribution}
To examine the impact of traffic distribution on the detection of the attack, we conducted experiments in three different topologies designed in Section~\ref{sec:emulation_setup}: \textbf{2ST}, \textbf{4ST} and \textbf{8ST}, where the distribution of traffic increases as the number of sub-trees increases in the topology of the network. Fig.~\ref{fig:allSubs} shows the classification results of SVM in different settings. As can be seen, the effectiveness of classification in \textbf{8ST} is significantly lower than \textbf{2ST}, and \textbf{4ST}, which is attributed to the fact that the former setting utilizes lower flow than the alternative settings during the attack scenarios. This low amount of traffic as compared with normal states of the network conceals the attack from the eyes of the detection approach. Further, the AUC for \textbf{2ST} and \textbf{4ST} is neck to neck with a small improvement in \textbf{4ST}. This is attributed to the fact that it benefits from more features as compared to \textbf{2ST}, i.e., $40$ features against $20$ features. Fig.~\ref{fig:allSubs:rf} depicts the classification results of RF model for different traffic distributions. As can be seen from the Figure, RF demonstrated similar behavior results as that for SVM. 

\subsubsection{The effect of features}
\begin{figure}
  \centering
  \includegraphics[width=0.90\linewidth]{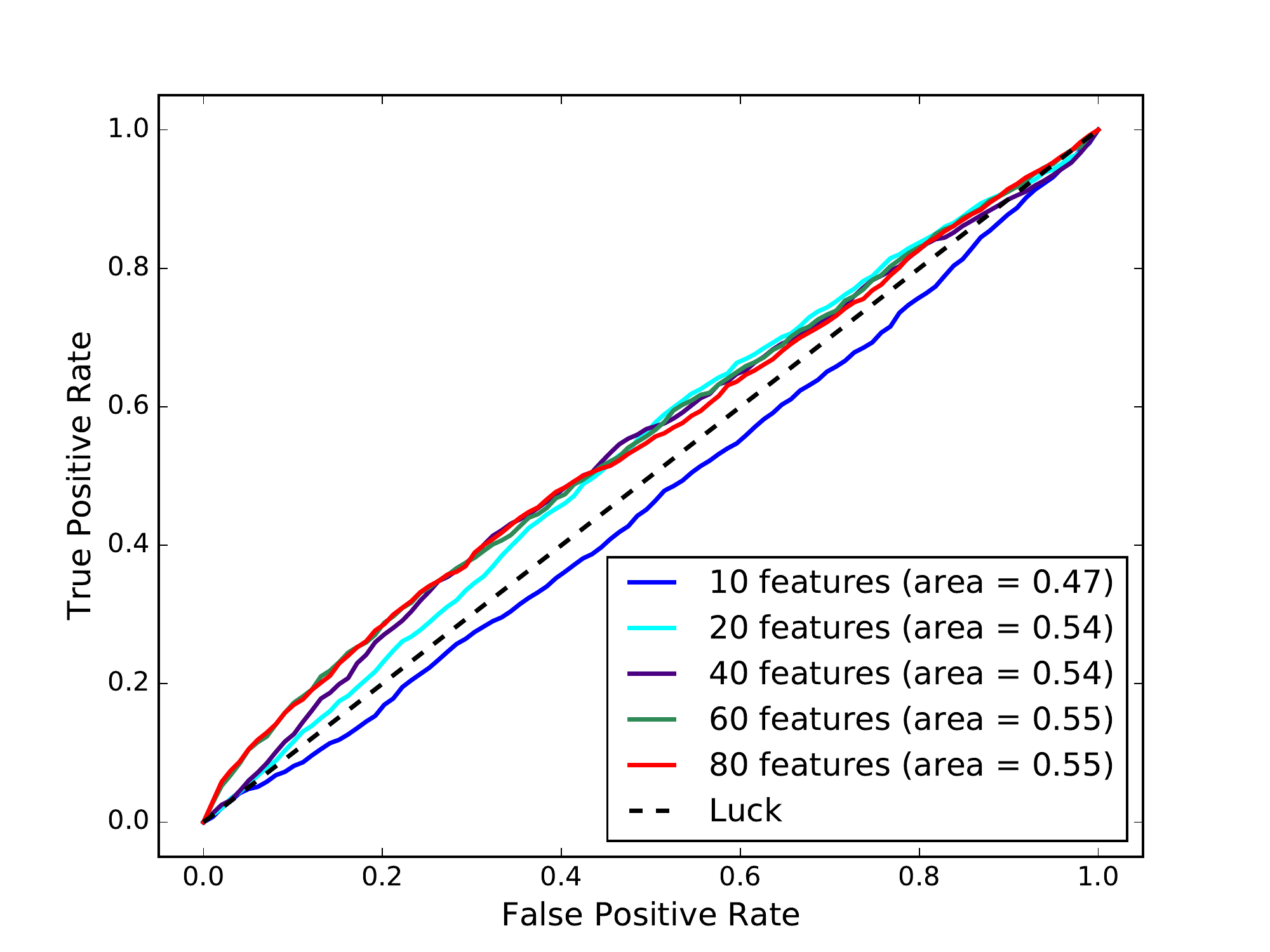}
  \caption{The effect of number of features on attack detection for 8-sub-tree.}
  \label{fig:8subFeatures}
\end{figure}

Prior studies in data mining have demonstrated that the performance of classification models highly depends on the selected features with regards to the classes. Further, a huge amount of data is required to be continuously processed as the network is a streaming and dynamic environment per se, which signify the importance of feature selection to reduce computational complexity. We hence vary the number of extracted features and evaluate the performance of classification algorithm in terms of AUC. Fig.~\ref{fig:4subFeatures} and Fig.~\ref{fig:8subFeatures} demonstrates the performance of classification for \textbf{4ST} and \textbf{8ST}, respectively. From Fig.~\ref{fig:4subFeatures}, we can see that the classification performance first indicates a positive correlation with the number of features and then saturates after an optimal value, i.e., $30$ numbers of features. This is an interesting results verifying that with a too small feature dimension we would fail to achieve the optimal performance. However, by only a limited number of features, we can achieve reasonable performance. This is important as in network environment we may access to a limited number of links %servers 
for feature extraction. 
Alternatively, Fig.~\ref{fig:8subFeatures} depicts the classification performance for the \textbf{8ST} setting. In contrast to \textbf{4St}, classification of \textbf{8St} setting is much lower. This indicates the importance of distribution of the Crossfire attack where with a enough distribution of attack standard machine learning approaches would fail to distinguish Crossfire attack traffic from background traffic. 

\subsubsection{The effect of network visibility}
\begin{figure}
    \centering   
    \includegraphics[width=0.90\linewidth]{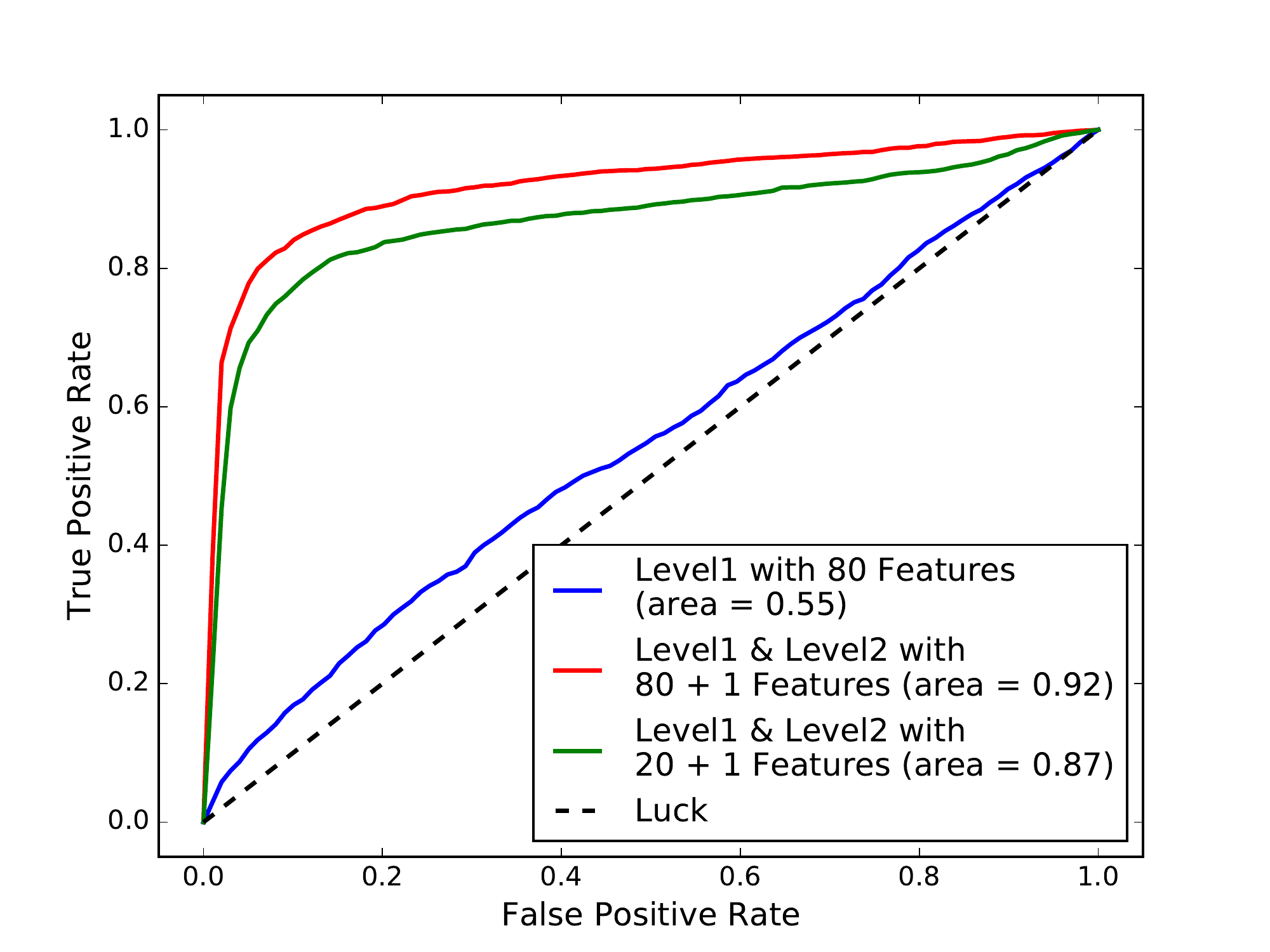}
  \caption{The effect of higher level features for SVM.}
  \label{fig:2level}
\end{figure}

\begin{figure}
  \centering
  \includegraphics[width=0.90\linewidth]{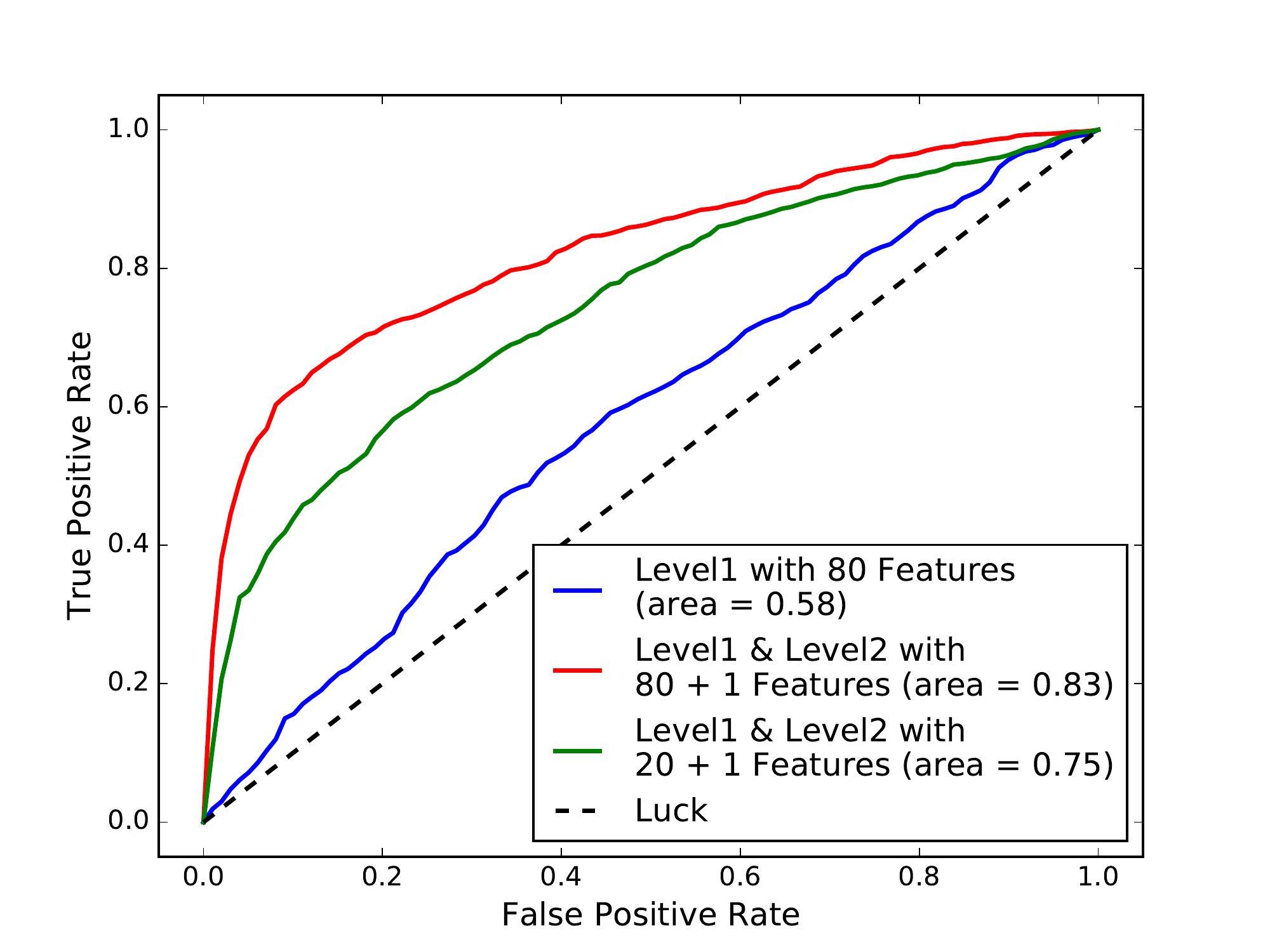}
  \caption{The effect of higher level features for RF.}
  \label{fig:2level:rf}
\end{figure}

Looking from the network aspect, an important factor for attack detection is the level of information we can gather about the traffic data of the network. To examine how features from different levels of the network affects the performance of traffic classification, we added the volume of one link from the upper level to the feature vector. More specifically, in \textbf{8ST} setting, we have the volume of $80$ decoy servers as a feature for the baseline. We also add the volume of a random link from one level upper to construct a $81$-dimension and $21$ -dimension feature vectors. Fig.~\ref{fig:2level} and Fig.~\ref{fig:2level:rf} demonstrate the performance of classification of traffic data in \textbf{8ST} setting for SVM and RF. Only adding one feature from the the upper level, even if there are less features from the lower level, improves the performance significantly, which highlights the importance of extracting features from different part of the network. %It is worth noting that the experimental results demonstrated while we need properly select the monitoring points, effective detection can be performed with a limited portion of features extracted from monitoring points. 

\section{Conclusion}
\label{sec:conclusion}

% Although the Crossfire attack is a potential threat to any network, in this paper, we show that it turns out that the adversary has also substantial obstacles in the successful attack execution, which exposes the attack to detection vulnerabilities. 
The Crossfire attack is considered to be one of the most difficult target-area link-flooding attacks to be detected. The attack uses a massively distributed large-scale botnet to generate multiple low-rate \emph{benign} traffic flows aiming to congest selected network link with the ultimate goal to disconnect the target area from the Internet. Although the Crossfire attack is a tremendous threat to any network, by analyzing the obtained data we show that the adversary has also substantial obstacles in the successful attack execution. As a result, this paper exposes detection vulnerabilities of the Crossfire attack by showing a correlation between coordination of the botnet traffic and the quality of the attack, and a correlation between the attack distribution and detectability of the attack. 

We also show that due to the bot synchronization there is a warm-up period after the attack is launched and before the target links are overwhelmed. Our results show that this period can be used for an early attack detection. %Therefore, for a detection mechanism, we suggest focusing on detecting any simultaneous sudden jump in traffic intensity at several links. %We are confident that our results can be used for the design of novel approaches for early Crossfire attack detection and attack mitigation measures.
In this paper a prototypical Crossfire attack detector is described, which exploits these vulnerabilities. For this, we utilize two supervised machine learning approaches: Support Vector Machine (SVM) and Random Forest (RF) for classification of network traffic to normal and abnormal traffic, i.e, attack traffic. In particular,  to show the feasibility of detection, we report on the trained scenarios using the link volume as the main feature set.
Finally, results of the attack detector are reported along with some future directions to improve the detector.

\section*{Acknowledgment}
This work is partially funded by the joint research programme UL/SnT-ILNAS on Digital Trust for Smart-ICT.

\balance
\bibliographystyle{ieeetr}
\bibliography{references}
 
\end{document}